\documentclass[prc,aps,nofootinbib,showkeys,showpacs,twocolumn,10pt,floatfix]{revtex4-1}
\usepackage{epsfig}
\usepackage{graphicx}
\usepackage{amsmath,amssymb,amsfonts}
\usepackage[small]{subfigure}
\usepackage[dvipsnames]{color}
\renewcommand{\d}{\mathrm{d}}                       
\newcommand{\E}{\mathcal{E}}                        

\newcommand{\fac}{(3\pi^2)}
\newcommand{\hw}{\hbar\omega}
\begin{document}
\title{Energy--density functionals inspired by effective--field theories:\\ Applications to neutron drops}
\author{J\'er\'emy Bonnard}\email{bonnard@ipno.in2p3.fr}
\affiliation{Institut de Physique Nucl{\'e}aire, IN2P3-CNRS, Universit{\'e} Paris-Sud,Universit{\'e} Paris-Saclay, F-91406 Orsay Cedex, France}

\author{Marcella Grasso}\email{grasso@ipno.in2p3.fr}
\affiliation{Institut de Physique Nucl{\'e}aire, IN2P3-CNRS, Universit{\'e} Paris-Sud,Universit{\'e} Paris-Saclay, F-91406 Orsay Cedex, France}

\author{Denis Lacroix}\email{lacroix@ipno.in2p3.fr}
\affiliation{Institut de Physique Nucl{\'e}aire, IN2P3-CNRS, Universit{\'e} Paris-Sud,Universit{\'e} Paris-Saclay, F-91406 Orsay Cedex, France}
\begin{abstract}
New energy--density functionals (EDFs) inspired by effective--field theories have been recently proposed.  The present work focuses on three such functionals, which were developed to produce satisfactory equations of state for nuclear matter. We aim to extend these functionals to treat finite systems including a spin--orbit contribution and pairing correlations. We illustrate here a first step towards this direction, namely a generalization of such functionals tailored to perform applications to neutron gases confined in harmonic traps. Sets of available \textit{ab initio} results are used as benchmark pseudo--data for adjusting the additional parameters (with respect to the nuclear matter case) that have to be introduced for finite--size systems. Several quantities are predicted and compared to \textit{ab initio} and other EDF results such as total energies, potentials, and density profiles. 
The associated effective masses are also analyzed. 
Two of these functionals globally provide predictions that are close to one another as well as to \textit{ab initio} values when available. It is shown that, in general, this is not the case for several currently used Skyrme functionals. Directions for improving the third functional are discussed.  
\end{abstract}
\date{\today}
\maketitle
%
\section{Introduction} \label{SecIntro}
Chiral effective--field theories (EFTs) provide a framework for building internucleon interactions that offers several advantages \cite{Bed02,EFTrev2,EFT0,EFT00}: (i) a direct link with QCD, (ii) consistent two--body and three--body forces, (iii) the possibility of systematic improvements by means of the order--by--order inclusion of diagrams, (iv) an estimate of theoretical uncertainties.
Such Hamiltonians are now commonly employed together with sophisticated many-body methods to perform \textit{ab initio} calculations for light nuclei or nuclear matter, see Refs.\cite{ab1,ab2,ab3,ab4,abrev0,abrev1,abrev2}. 

Whereas reliable \textit{ab initio} calculations are limited to a small number of particles, EDF theories \cite{EDF,Sto07} represent the unique approach allowing us to investigate the nuclear chart as a whole as well as dense matter, traditionally on the basis of phenomenological effective interactions such as Skyrme and Gogny forces \cite{sk0,sk1,sk2,gogny,gogny2}. 

Recently, efforts have been undertaken to bridge EFT and EDF theories \cite{EFTEDF0,EFTEDF}. 
The aim in borrowing concepts from EFT is the development of a new generation of functionals potentially able to encode beyond--mean--field effects, to describe correlated exotic nuclei, and, more importantly, to involve less adjustable parameters. The construction of next--to--leading order Skyrme--like effective forces addressing regularization and renormalizability issues \cite{O2Sk1,O2Sk2} and a first attempt for defining power--counting schemes to build EDFs \cite{NLOEDF} are examples of steps towards this direction. Note that, in parallel, alternative extensions of the Skyrme EDF that also incorporate higher--order contributions (but not in the sense of the Dyson perturbative expansion) without invoking EFT techniques have been proposed \cite{SN2LO}.

The present work focuses on three EFT--inspired functionals, namely YGLO\footnote{Yang Grasso Lacroix Orsay.} \cite{YGLO}, KIDS\footnote{Korea IBS Daegu Sungkyunkwan.} \cite{KIDS1}, and ELYO\footnote{Extended Lee-Yang Orsay.} \cite{ELYO}. YGLO consists of a hybrid EDF gathering standard Skyrme--type velocity-- and density--dependent contributions together with a resummed term whose formal expression is based on a resummation formula used in EFTs for systems with large scattering lengths \cite{EFT2,EFT2b}. The KIDS functional is written as a power expansion in the Fermi momentum with the same first orders as those naturally emerging in EFTs. Finally, ELYO relies on the equation of state (EOS) of very dilute neutron matter (first obtained by Lee and Yang in the 1950s \cite{LY,Bis73} and derived more recently within the framework of EFTs \cite{EFT1}) extending its validity domain to reach density regimes of interest for finite nuclei via the introduction of a density--dependent neutron--neutron scattering length.

The aforementioned studies mainly concern nuclear matter for which the link between EFT and EDF theories is easier.  Whereas KIDS was already applied to atomic nuclei \cite{KIDS2,KIDS3,KIDS4}, this is not the case for the YGLO and ELYO functionals that are therefore characterized only by the EOS to which they lead. In this paper, our purpose is to generalize these EDFs to enable the treatment of finite systems. In particular, we address systems composed exclusively of neutrons confined in isotropic harmonic traps. These drops offer a simple model for extremely neutron--rich nuclei (where the unbound valence neutrons are trapped by an external well from the core).  Owing to their isospin composition, these systems allow for a direct assessment of the isovector channel of the functionals: This is indeed the less constrained part in common EDFs, which are  usually fitted to measured observables of stable or near--stable isotopes and, hence, produce large dispersions in the predictions for strongly isospin--asymmetric nuclei. Moreover, the properties of low--density neutron systems are crucial ingredients for understanding systems located in the inner crust of neutron stars. Last, a strong linear correlation has been discovered between the radii of neutron drops and the neutron skin thickness of ${}^{208}$Pb and ${}^{48}$Ca \cite{skin}.

All these features of neutron drops have motivated a large number of theoretical studies, especially based on \textit{ab initio} calculations \cite{NCSMQMC,QMC,NCSMEFT}.
Contrary to infinite matter, the spin--orbit (and the tensor, omitted here) interaction impacts the properties of finite--size systems such as their shell structure. Furthermore, the role played by superfluidity in affecting the shell structure  cannot be neglected. 
The {\it{ab initio}} description of these properties provides precious pseudo--data to benchmark the nuclear EDF approach 
and to constrain functionals in specific spin-isospin channels that can hardly be optimized otherwise \cite{QMC,UNEDF,Pudliner,Smerzi,Drut}. 
In particular, we resort here to a collection of available \textit{ab initio} results to adjust the extra parameters needed to extend the YGLO, KIDS, and ELYO functionals to finite systems. In the present work, we generalize the two functionals YGLO and ELYO to account for spin--orbit and pairing contributions. In addition, we discuss the constraints on the 
effective masses induced by the chosen strategies for extending the functionals.  
The available applications of the KIDS EDF to doubly magic nuclei include the spin--orbit interaction \cite{KIDS4}. This functional is complemented here by explicitly treating pairing correlations.  

The paper is organized as follows.  Section \ref{SecEFTfunc} reviews the properties of the three EFT--guided EDFs considered in this work.  Their generalization for applications to neutron drops is detailed in Sect. \ref{SecDrop}.  Results are presented and discussed in Sect. \ref{SecRes} and conclusions are drawn in Sect. V.  Some expressions for the Skyrme EDF, which will be useful through this paper, are given in Appendix A. 
%
%
\section{EFT-inspired functionals} \label{SecEFTfunc}
\begin{figure}[htbp]
\begin{center}
\includegraphics[scale=0.9,width=\columnwidth]{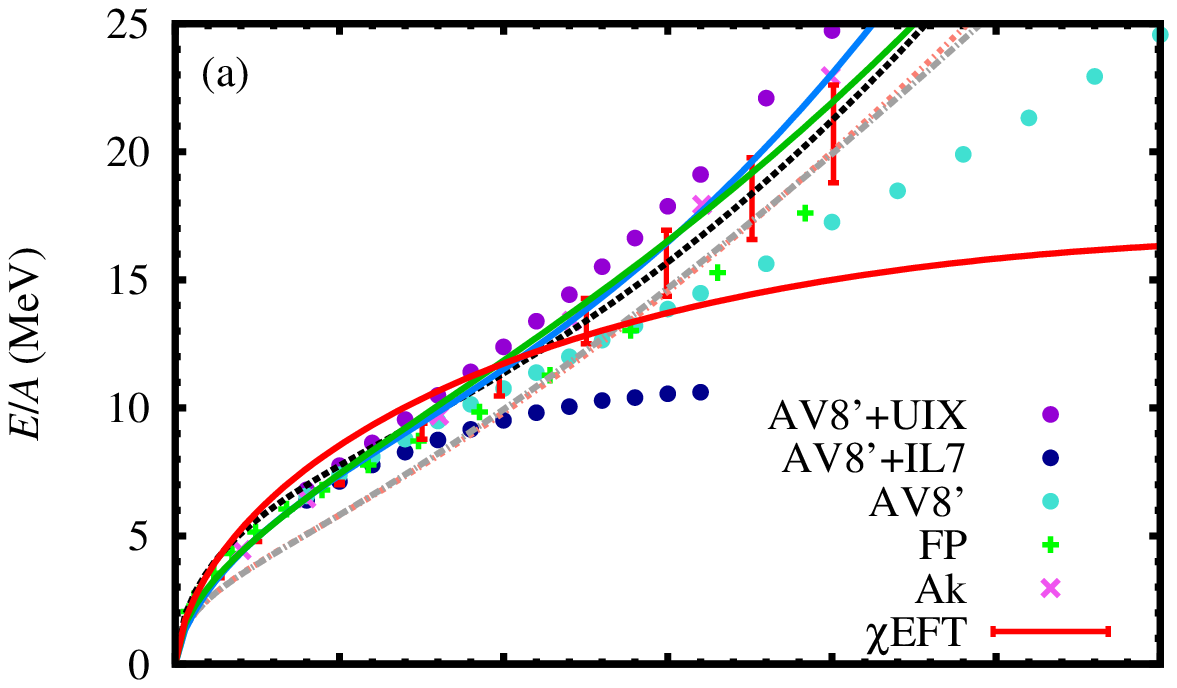} \\
\vspace{-1.65cm}
\includegraphics[scale=0.9,width=\columnwidth]{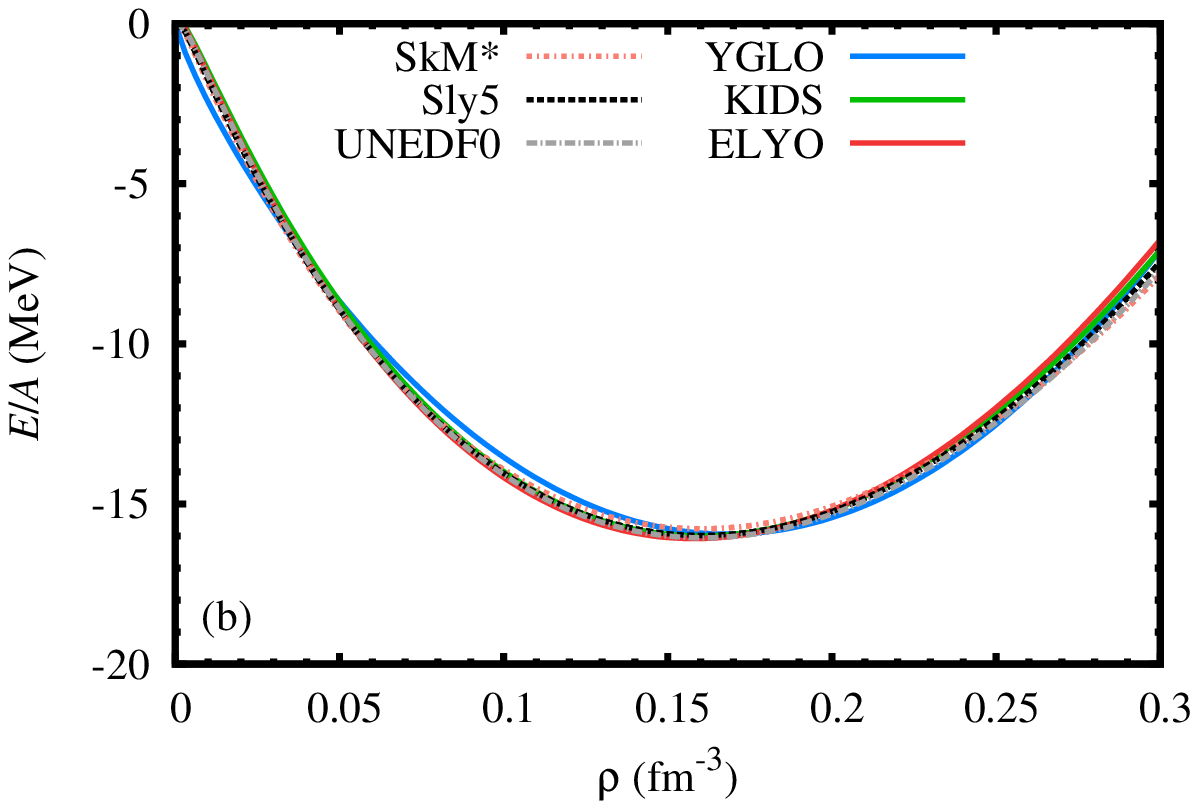}
\caption{EOSs of (a) PNM and (b) SNM obtained with the three EFT--inspired functionals (full lines) compared with those provided by some commonly used Skyrme EDFs (dashed lines), SkM* \cite{SKM}, Sly5 \cite{SLY5}, UNEDF0 \cite{UNEDF0}. For neutron matter, also shown are the \textit{ab initio} results from: QMC calculations \cite{NCSMQMC,QMC} using the AV8' two--body force only \cite{AV8} or using AV8' supplemented by the UIX \cite{UIX} or IL7 \cite{IL7} three--body interactions (dots); Friedman-Pandharipande \cite{QMCFP} (FP, green plus); Akmal \textit{et al}. \cite{QMCAkmal} (Ak, magenta crosses); and Ref. \cite{DSS} ($\chi$EFT, red bars). For the latters that are based on chiral EFT, the size of the bars represents the theoretical uncertainties.}
\label{fig_EoSMatter}
\end{center}
\end{figure}

We briefly describe here the  three functionals in their original versions, that is, as they were proposed for nuclear matter. We therefore focus on the corresponding EOSs, plotted in Fig. \ref{fig_EoSMatter}. For the YGLO EDF, we represent here only the YGLO (Akmal) case \cite{YGLO}, which we call for simplicity YGLO.  
In this figure, we display some EOSs obtained with selected conventional sets of Skyrme parameters. We choose as an illustration the SkM* \cite{SKM}, the Sly5 \cite{SLY5}, and the UNEDF0 \cite{UNEDF0} parametrizations. The parameters of Sly5 have been specifically adjusted to reproduce the EOS in neutron matter while UNEDF0 is one of the latest adjusted Skyrme functionals. For comparison, we report the EOSs for neutron matter [Fig \ref{fig_EoSMatter}(a)] from QMC calculations \cite{NCSMQMC,QMC} using the AV8' two--body force only \cite{AV8} or using AV8' supplemented by the UIX \cite{UIX} or IL7 \cite{IL7} three--body interactions. The Friedman-Pandharipande (FP) results of Ref. \cite{QMCFP} and the Akmal \textit{et al}. (Ak) results of Ref. \cite{QMCAkmal} are also shown.

All Skyrme and EFT--inspired functionals lead to a similar behavior for the EOS of symmetric nuclear matter (SNM) and, qualitatively, to the same trend in pure neutron matter (PNM), at least for densities up to $\sim$0.1 fm$^{-3}$. We emphasize that, compared to the Skyrme case where the number of adjustable parameters for reproducing the EOS of matter is nine, the YGLO functional has seven parameters and the ELYO functional only five. In this sense, the ELYO results can indeed be considered very satisfactory. In spite of the strongly reduced number of parameters (almost one--half compared to the Skyrme case), the EOS produced for SNM is very good and the EOS predicted for PNM is satisfactory up to neutron densities at play in finite nuclei. The rest of the ELYO PNM EOS (at higher densities) is comparable to PNM EOSs provided by other Skyrme functionals such as SIII \cite{siii} or SkP \cite{RefSk} as shown in Ref. \cite{ELYO}.   

The KIDS and YGLO PNM EOSs are almost identical for densities $\rho < 0.15 \text{ fm}^{-3}$. The Sly5 PNM EOS follows also closely these two EOSs with some departure at low densities $\rho< 0.05\text{ fm}^{-3}$. The UNEDF0 and SkM* PNM EOSs are located at lower energies (compared to Sly5, YGLO, and KIDS) over the whole range of densities displayed in the figure. 
These differences will be useful in some cases for discussing to what extent the properties of neutron drops are sensitive to infinite matter EOSs. 

%
%
\subsection{The YGLO functional}
The YGLO functional \cite{YGLO} produces the following EOS, 
\begin{equation} \label{YGLOEoS}
   \dfrac{E}{A} = K_\beta(\rho) + Y_\beta(\rho) \rho + D_\beta \rho^{5/3} + F_\beta \rho^{1+\alpha},
\end{equation}
with $\alpha=0.7$ and where $\rho$ is the total density of either PNM ($\beta=0$) or SNM ($\beta=1$). 
$K_\beta$ stands for the usual kinetic contribution and $Y_\beta(\rho)$ for a resummed term of the form
\begin{equation}
   Y_\beta(\rho) = \dfrac{B_\beta}{1-R_\beta\rho^{1/3}+C_\beta\rho^{2/3}}.
\end{equation}
$B_\beta$ and $R_\beta$ are constrained by imposing the correct limit at very low density, that is by matching the resummed term with the Lee-Yang expansion \cite{LY}
up to second order in $(a_s k_F)$ where $a_s$ is the $s$--wave scattering length and $k_F$ is the Fermi momentum. 
They are hence expressed as
\begin{equation} \label{YGLOld}
   \begin{split}
   & B_\beta = \dfrac{2\pi\hbar^2}{m}\dfrac{\nu-1}{\nu}a_s, \\
   & R_\beta = \dfrac{6}{35\pi}\left(\dfrac{6\pi^2}{\nu}\right)^{1/3}(11-2\ln2)a_s.
   \end{split}
\end{equation}
$\nu=2$ (4) is the degeneracy for $\beta=0$ (1) and $m$ is the nucleon mass taken to be equal for protons and neutrons. Note that, in Eq. \eqref{YGLOld}, different values of $a_s$ are employed for PNM and SNM.
The $D_\beta$, $F_\beta$, and $C_\beta$ parameters for $\beta=0$ and $1$ were obtained by a fit of PNM and SNM EOSs, Eq. \eqref{YGLOEoS}, on the two sets of quantum Monte-Carlo (QMC) pseudo--data taken from Refs. \cite{QMCGez,QMCFP} and \cite{QMCGez,QMCAkmal}, yielding two possible parametrizations called YGLO (FP) and YGLO (Akmal), respectively. In the present work, we only consider the latter (denoted simply by YGLO) but all the drawn conclusions also apply to the former.

It is interesting to mention that another functional based on a resummed formula was suggested to reproduce the unitary limit of Fermi gases and neutron matter at low density \cite{lacroix1,lacroix2,UG}.
%
%
\subsection{The KIDS functional}
%
The KIDS functional \cite{KIDS1} consists in a power expansion in the Fermi momentum $k_F=(6\pi^2\rho/\nu)^{1/3}$ equivalent to
\begin{equation}
   \dfrac{E}{A} = K_\beta(\rho) + \sum_{i=0}^3 C_\beta^{(i)} \rho^{1+i/3}. \label{eq:kids}
\end{equation}
Here, we consider the specific ad-2 parametrization on which is based the application to nuclei \cite{KIDS2,KIDS3,KIDS4} and that does not retain the logarithmic term. The $C_\beta^{(i)}$ coefficients are determined by a fit on SNM properties at saturation density and QMC calculations for PNM. 
%
%
\subsection{The ELYO functional}\label{Sec_ELYOMatter}
%
The ELYO functional \cite{ELYO} is designed to provide an EOS for PNM corresponding to the first terms of the Lee-Yang formula with only $s$--wave contributions. It is constructed in such a way 
that the Lee-Yang--type formula holds at all density scales for neutron matter. The resulting EOSs for both SNM and PNM 
may be written as pure $s$--wave Skyrme--like EOSs, that is neglecting the $p$--wave term ($\E_2=0$ in Appendix A), and with the power of the density--dependent term equal to 1/3. From the EOS of PNM, the Skyrme parameters are linked to the low--energy constants through
\begin{equation} \label{condLY}
\begin{split}
  &t_0(1-x_0)=\dfrac{4\pi\hbar^2}{m}a_s ,\\
  &t_1(1-x_1)=\dfrac{2\pi\hbar^2}{m} (r_sa_s^2+0.19\pi a_s^3), \\
  &t_3(1-x_3)=\dfrac{144\hbar^2}{35m} \fac^{1/3} (11-2\ln 2)a_s^2,
\end{split}
\end{equation} 
where $a_s=-18.9$ fm is the neutron--neutron scattering length and $r_s=2.75$ fm is the associated effective range. 
The $t_i$'s coefficients are adjusted to generate a satisfactory EOS for SNM around the equilibrium point, whereas the $x_i$'s parameters are given by Eq. \eqref{condLY}. This implies that the PNM EOS does not depend on the adjusted parameters. 
However, such a direct mapping to the Lee-Yang formula is valid only when $|a_sk_F|\leq 1$, that is in the very low-density regime up to $10^{-6}$ fm$^{-3}$. To allow using the low--density expansion at all density scales, the constraint $|a_sk_F|\leq 1$ is extended by assuming a density--dependent $a_s$:
\begin{gather} \label{as_rho}
a_s(\rho)=
\left\lbrace
\begin{array}{ll}
-18.9 \text{ fm} & \text{if }   (18.9k_F)\leq \Lambda \\
\\
- \Lambda/(3\pi^2\rho)^{1/3}  & \text{if } (18.9k_F)> \Lambda
\end{array}\right.,
\end{gather}
with $\Lambda\leq 1$ a chosen limit value for $|a_s(\rho)k_F|$.
Furthermore, the effective range in the regime where $a_s$ departs from its bare value is used as an adjustable parameter and $r_s=-4.5$ fm was found to give a reasonable PNM EOS for $\Lambda=1$ at least up to densities of interest for finite nuclei (see Fig. \ref{fig_EoSMatter}).
The set of parameters $\{ x_i \}$ are thus tuned by the density--dependent neutron--neutron scattering length.

For all the above--described functionals, the different EOSs for intermediate asymmetries may be deduced via the so--called  parabolic approximation where the symmetry energy is computed as the difference between the EOSs of PNM and SNM.
%
%
%
%
\section{Extension to neutron drops} \label{SecDrop}
It is well known that the adjustment of a functional done only on infinite matter is not enough to correctly describe finite systems. 
This is the reason why, in general, additional constraints on specific nuclei are added in the fitting process. As discussed in Sec. \ref{SecIntro}, 
the spin--orbit interaction should be added to properly account for shell effects. In addition, pairing correlations should be explicitly incorporated within the EDF. In the present section, we describe for each functional how these new components are introduced. For YGLO and ELYO, guided by the procedure employed for KIDS, we also propose a strategy to separate the functional into density--dependent and velocity--dependent terms (which generate an effective mass in leading--order calculations).     
%
\begin{figure}[htbp]
\begin{center}
\includegraphics[width=\columnwidth]{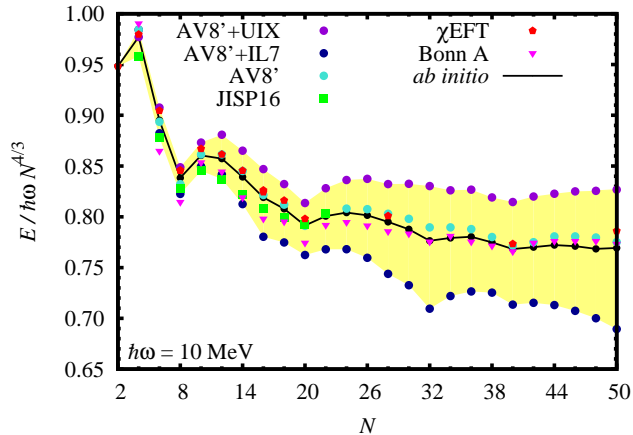}
\caption{Energies of neutron drops in a $\hw=10$ MeV trap, scaled by $\hw N^{4/3}$, obtained from a variety of \textit{ab initio} calculations with different two-- and three--body interactions. The dots refer to the three QMC calculations producing the EOSs plotted in Fig.  \ref{fig_EoSMatter}(a). The squares indicate results from a configuration--interaction method \cite{NCSMQMC} with the JISP16 force \cite{JISP16}. The pentagons represent no--core shell--model and coupled--cluster calculations with an interaction derived from chiral EFT \cite{NCSMEFT}. The average of all these results is denoted as ``\textit{ab initio}" (full line). Also shown (but not included in the reference data set) are relativistic Brueckner-Hartree-Fock calculations \cite{tensor,tensor2} with the Bonn A interaction \cite{BonnA} (triangles). The upper limit for free neutrons gives an horizontal line at $E/\hw N^{4/3} \approx 1.082$.}
\label{fig_abinitio}
\end{center}
\end{figure}
\subsection{EDF treatment and adjustment procedure} \label{EDFforDrops}
\subsubsection{Generalities}

Let us now consider a system composed of a finite number $N$ of neutrons trapped in an isotropic harmonic well of frequency $\omega$. Such a potential makes relevant the use of spherical coordinates. Within the framework of EDF theory, the total energy of the system is given by $E= \int \d^3\vec{r} \E(\vec{r})$ where the energy density $\E(\vec{r})$ is decomposed in terms depending on the neutron $\rho$, kinetic $\tau$, spin--current $\vec{J}$ , and anomalous pairing $\tilde{\rho}$ densities as
\begin{equation} 
   \E(\vec{r}) = \mathcal{T}(\vec{r}) +\E_\mathrm{\omega}(\vec{r})+ \E_\mathrm{c}(\vec{r}) + \E_\mathrm{so}(\vec{r}) + \E_\mathrm{pp}(\vec{r}).  \label{edffull}
\end{equation}
$\mathcal{T}(\vec{r})=\hbar^2\tau(\vec{r})/2m$ is the kinetic contribution with a neutron mass taken as $\hbar^2/m=41.44$ MeVfm$^2$. $\E_\mathrm{\omega}(\vec{r})$ describes the trap contribution related to the potential $m\omega^2\vec{r}^{\,2}/2$. 
The drop being localized, there is no center-of-mass correction. In Eq. \eqref{edffull}, we explicitly separate the spin--orbit $\E_\mathrm{so}(\vec{r})$ and the pairing $\E_\mathrm{pp}(\vec{r})$ contributions from the rest that is generically denoted by $\E_\mathrm{c}$.  
We adopt a mixed surface/volume pairing interaction, 
\begin{equation}
   V(\vec{r}) = V_\mathrm{pp} \left( 1-\dfrac{1}{2}\dfrac{\rho(\vec{r})}{\rho_c}\right)\delta(\vec{r}),
\end{equation}
with $\rho_c=0.16$ fm${}^{-3}$ and a standard smooth (diffuseness of 1 MeV) cut-off at 60 MeV in the quasiparticle spectrum.

The quasiparticle wave functions from which the various densities are built are obtained self--consistently by solving Hartree-Fock-Bogoliubov (HFB) equations. The expressions of the densities in spherical symmetry, as well as the particle-hole and particle-particle fields in the case of a Skyrme EDF, may be found in Ref. \cite{RefSk}. For the present purpose we updated the spherical HFB code HFBrad \cite{hfbrad} to be able to treat finite--size systems such as neutron drops with the functionals YGLO, KIDS, and ELYO. The results reported in this work have been computed with a radial space coordinate discretized in 150 steps of 0.2 fm and by taking into account orbitals up to angular momentum $j=15/2$. 
These values are such that for all the systems treated here the calculations are well converged.
The other numerical parameters are those defined by default in the program \cite{hfbrad}.

Adapting to finite neutron droplets the EFT--inspired functionals defined in Sect. \ref{SecEFTfunc} requires to establish a possible expression for their $\E_\mathrm{c}$ part in terms of the densities. To this end, we rely on the traditional Skyrme energy functional:  When possible, the terms of the YGLO, KIDS, and ELYO EOSs  are identified as stemming from Skyrme--like terms of the functional $\E_i$ ($i=0,1,2,3$, see Appendix A). 
It is worth mentioning that, as in the Sly5 parametrization, the so--called $J^2$ contributions are not neglected.
The resummed part of YGLO, for which a mapping with usual Skyrme terms is not possible, is directly transposed by extending functions to functionals: $Y(\rho)\rightarrow Y[\rho(\vec{r})]$. This procedure necessitates the introduction of new parameters that will be adjusted on a set of pseudo--data extracted from \textit{ab initio} calculations (see Sec. \ref{sec:IIB}). 

\subsubsection{Selection of \textit{ab initio} pseudo--data and fitting protocol}

For a given neutron number, we choose as reference pseudo--data the average of the  \textit{ab initio} energies for $\hw=10$ MeV compiled in Fig. \ref{fig_abinitio}. In the following, adjustments of the functionals are performed on this average. Figure \ref{fig_abinitio} displays in particular QMC calculations 
producing the EOSs plotted in Fig.  \ref{fig_EoSMatter}(a), configuration--interaction calculations \cite{NCSMQMC} with the JISP16 force \cite{JISP16}, no--core shell--model and coupled--cluster calculations with an interaction built within chiral EFT \cite{NCSMEFT}. The average of all these results is denoted as ``\textit{ab initio}" (full line). Figure \ref{fig_abinitio} also shows relativistic Brueckner-Hartree-Fock calculations \cite{tensor,tensor2} with the Bonn A interaction \cite{BonnA}.
The dispersion of the different estimates observed when the neutron number increases, represented by the yellow area, is well understood from the properties of the corresponding interactions, as discussed in details in Refs. \cite{NCSMQMC,tensor}. The reference values within this area seem physically reasonable since they quantitatively agree well with $\chi$EFT \cite{NCSMEFT} and Bonn A \cite{tensor,tensor2} results (for $N\geq 30$), the latter not being comprised in the benchmark data set. Note that, as a consequence, adjusting the EDFs to calculations relying on EFT interactions only, in the spirit of EFT--inspired functionals, leads to parameters very close to those obtained by fitting to the average.

The new parameters entering in $\E_c$ as well as the spin--orbit coupling and pairing strength, $V_\mathrm{so}$ and $V_\mathrm{pp}$ respectively, are determined simultaneously by adjusting HFB results for the scaled energy $E/\hw N^{4/3}$ on the benchmark data for $N = 8,12,14,16,$ and 20 with $\hw=10$ MeV. These neutron numbers are retained for the fit insofar as they belong to the range where the various approaches agree rather well thus ensuring well constrained reference data. Indeed, $N\approx 30$, for instance, could in principle also be a good constraint for $V_\mathrm{pp}$ (half--filled $fp$--shell systems). Nevertheless, as observed on Fig. \ref{fig_abinitio}, the benchmark data exhibit a surprising shell closure at $N=32$ that is not due to the harmonic trap and that mainly comes from the AV8'+IL7 calculation. This closure artificially reduces the reference energy of the $N\approx 30$ drops. Once the parameters are adjusted, the physical quantities computed for $N> 20$ with $\hw=10$ MeV and for any value of $N$ with other trap frequencies will allow us to assess the quality of the resulting functionals and can be regarded as predictions.

In the following subsections, the $\E_\mathrm{c}$ functional derived for the considered EDFs is expressed as holding for nuclei for generality.
Thus, both proton and neutron densities will appear in the expressions, respectively labeled by $p$ and $n$ indices.

\subsection{Discussion on the KIDS functional}
%
The strategy to apply the KIDS EDF to finite nuclei has been described in Refs. \cite{KIDS2,KIDS3,KIDS4} and we only provide here details relevant for the forthcoming discussion. The first step is to match Eq. (\ref{eq:kids}) into a Skyrme--like EOS provided by a functional of the type
\begin{equation} \label{KID1}
   \E_c=  \E_0 + \E_1 + \E_2 + \E_3 + \E_{3'}+  \E_{3''}.
\end{equation} 
$\E_0$ is related to the $ C_\beta^{(0)}$ term with $t_0$ and $x_0$ given by the relations
\begin{eqnarray}
\frac{3}{8}t_0 = C^{(0)}_1, ~~\frac{1}{4} t_0 (1-x_0)=  C^{(0)}_0. \nonumber
\end{eqnarray}
The two terms for $i=1$ and $i=3$ that correspond respectively to the $\rho^{4/3}$ and $\rho^2$ parts in the EOS \eqref{eq:kids} are directly interpreted as distinct density--dependent terms, associated to $\E_3$ and $\E_{3''}$ in Eq. (\ref{KID1}) with $\alpha=1/3$ and $\alpha''=1$, respectively. In a similar way as for $\E_0$, one may easily relate the $(t_3, x_3)$ and $(t_{3''}, x_{3''})$ coefficients to the values of $C^{(1)}_\beta$ and $C^{(3)}_\beta$.
 
The term for $i=2$ in the EOS offers more flexibility because of its $\rho^{5/3}$ density dependence. It could indeed be interpreted either as a contribution from $(\E_1+ \E_2)$, or as a third density--dependent term $\E_{3'}$ with $\alpha'=2/3$. In Refs. \cite{KIDS2,KIDS3}, an extra parameter $W$ was introduced to share the $i=2$ term between these two contributions: $W$ denotes the fraction in the $i=2$ term coming from $(\E_1 + \E_2)$ whereas $(1-W)$ denotes the $\E_{3'}$ contribution. Assuming further for simplicity that $x_1=x_2 = 0$, the coefficients $t_1$, $t_2$, $t_{3'}$ and $x_{3'}$ can be expressed as a function of $C^{(2)}_0$, $ C^{(2)}_1$, and $W$. 
Similar strategies are followed for the two other functionals considered in the present work.  
Note that, except for the pairing strength, we use for the KIDS functional the values of parameters inferred by fitting properties of closed-shell nuclei in Refs. \cite{KIDS2,KIDS3} including the spin--orbit contribution.   

%
\subsection{Extension of the YGLO functional}
\label{sec:IIB}
Comparing the YGLO EOS given by Eq. (\ref{YGLOEoS}) with standard Skyrme EOSs, one can establish a correspondence between some terms of the underlying functionals, and accordingly define the expression of its central part.
Thus, the $F_\beta$ term may be written as a density--dependent contribution $\E_3$ with $\alpha=0.7$ provided that
\begin{equation} \label{t3YG}
\begin{split}
   &t_3=16F_1, \\
   &t_3(1-x_3)=24F_0.
\end{split}
\end{equation}
The $D_\beta$ term demands more care in the sense that a ${5/3}$ power of $\rho$ may originate from a velocity-dependent term ($\E_{1,2}$), from a density--dependent one ($\E_{3'}$ with $\alpha'=2/3$), or from any combination of both. Consequently, the identification leads to two equations for six unknown parameters ($t_i$, $x_i$ for $i=1,2,3'$). To remove this ambiguity, we follow Refs. \cite{KIDS2,KIDS3} and introduce a new coefficient $W$ that governs the proportion $D^{(12)}_\beta$ of $D_\beta$ coming from a velocity--dependent term so that 
\begin{equation} \label{splitex}
   D^{(12)}_\beta= WD_\beta,\quad D^{(3')}_\beta=  (1-W)D_\beta, 
\end{equation}
with $D^{(3')}_\beta$ the part corresponding to a density--dependent term. As highlighted by Eq. \eqref{splitex}, $W$ weights the contribution related to the effective mass without modifying the EOS. It allows us to fully determine the coefficients of the density--dependent contribution $\E_{3'}$, 
\begin{equation} \label{t32YG}
\begin{split}
   &t'_3=16(1-W)D_1, \\
   &t'_3(1-x'_3)=24(1-W)D_0.
\end{split}
\end{equation}
Regarding the $\E_{1,2}$ part, we now have four parameters and two equations. At this stage, several strategies may be adopted. First, we tried to follow 
the same prescription as in Refs. \cite{KIDS2,KIDS3} and imposed $x_1=x_2=0$. However, we found that the numerical solutions of the HFB equations become unstable for $|W|>0.2$. On the other side, for $|W|<0.2$, the quality of the obtained fit is not acceptable.
The reason why such an approach fails for the YGLO functional whereas it works well in the KIDS case may be explained by comparing the resulting $t_{1,2}$ values: Contrarily to the KIDS functional (and to usual Skyrme EDFs), the $t_1$ and $t_2$ parameters have the same sign for any value of $W$ in the YGLO case. 

An alternative strategy consists in retaining only $s$--wave terms, that is $t_2=x_2=0$, as in the case of the ELYO EDF, which yields
\begin{equation} \label{t1YG2}
\begin{split}
  & t_1=\dfrac{80}{9}W \left(\dfrac{3\pi^2}{2}\right)^{-2/3}D_1, \\
  & t_1(1-x_1)=\dfrac{40}{3}W (3\pi^2)^{-2/3}D_0.
\end{split}
\end{equation}
Finally, we end up with an YGLO functional written as 

\begin{equation} \label{YGLO2}
\begin{split}
   \E_c=&\bigl(2Y_1[\rho]-Y_0[\rho]\bigl)\rho^2 -2\bigl(Y_1[\rho]-Y_0[\rho]\bigr)\bigl(\rho_n^2+\rho_p^2\bigr) \\
       &+ \E_1 + \E_3 + \E_{3'}, 
       \end{split}
\end{equation}
with parameters given by Eqs. (\ref{t3YG}), (\ref{t32YG}), and (\ref{t1YG2}). This strategy is applied in the following.

%
\subsection{Extension of the ELYO functional}
We now consider the ELYO functional and follow a similar approach. In this case, the form of the central part is easier to interpret since it is defined as a pure $s$--wave Skyrme--like functional with parameters given by Eq. \eqref{condLY} and $\alpha=1/3$. For finite systems such as neutron drops, the scattering length becomes a functional $a_s[\rho(\vec{r})]$ of the total density. The EDF may thus be written as a Skyrme one with $x_i$ parameters depending on $\rho(\vec{r})$, or may equivalently be recast as
\begin{equation} \label{ELYO1} 
\begin{split}  
 &\E_c = \E^\mathrm{Sk}_c - \biggl[X_0 a_s[\rho] + X_3 \rho^{\alpha} a_s^2[\rho] \biggr] \biggl[\dfrac{1}{2}\rho^2 
 - \sum_{q=n,p}\!\!\rho_q^2 \biggr]  \\
 &- X_1 B_s[\rho]  \\
    &\!\!\!\!\times\biggl[\dfrac{1}{2}\rho\tau+\dfrac{3}{8}(\vec{\nabla}\rho)^2 - \dfrac{1}{4}\vec{J}^{\,2} - \sum_{q=n,p}\!\!\Big(\rho_q\tau_q+\dfrac{3}{4} (\vec{\nabla}\rho_q)^2\Bigr) \biggr], 
\end{split}
\end{equation}
with
\begin{align*}
  & X_0 =\dfrac{2\pi\hbar^2}{m}, \\
  & X_3 =\dfrac{12\hbar^2}{35m}(11-2\ln 2)(3\pi^2)^{1/3}, \\
  & X_1 =\dfrac{\pi\hbar^2}{2m},
\end{align*}
and where $a_s[\rho]$ is taken from Eq. \eqref{as_rho} [the number density is replaced by the local part $\rho(\vec{r})$ of the one-body density matrix]. $\E^\mathrm{Sk}_c$ stands for a usual ($s$--wave) Skyrme EDF with parameters $t_{0,1,3}$ from the fit on SNM properties of Ref. \cite{ELYO} and $x_{0,1,3}=1$. For compactness, we have introduced the notation 
\begin{eqnarray}
B_s [\rho] & \equiv & \biggl[ r_s a_s^2[\rho] +0.19\pi a_s^3[\rho]\biggr].
\end{eqnarray}

One observes that for neutron drops as for PNM the EDF does not depend on the phenomenological parameters $t_i$ as they only enter in $\E^\mathrm{Sk}_c$ that cancels out in that case. The value of the effective range $r_s$ also depends on the density and, by extension, on the position $\vec{r}$:  $r_s=2.75$ fm wherever the density is small enough so that the scattering length is equal to $-18.9$ fm; $r_s=-4.5$ fm for higher densities where $a_s$ becomes a functional of $\rho(\vec{r})$.

Equation \eqref{ELYO1} relies on the hypothesis that the terms of the Lee-Yang expansion match one-by-one with those of the Skyrme EOS. More precisely, it is assumed that each of the $k_F^4$ ($\rho^{4/3}$) and $k_F^5$ ($\rho^{5/3}$) powers respectively identifies to a density--dependent $\E_3$ and to a velocity--dependent $\E_1$ contribution. Nevertheless, it is possible to split the $\rho^{5/3}$ term into $\E_1$ plus an additional density--dependent term $\E_{3'}$ corresponding to $\alpha'=2/3$ (by resorting once again to a new parameter $W$).  Actually, this step has turned out to be necessary in practice to get a satisfactory fit. Without this splitting, a reasonable fit could be obtained only if the value of $r_s$ or $\Lambda$ were strongly modified, thus entailing a severe degradation in the PNM EOS.
Due to the introduction of the parameter $W$, the second line in Eq. \eqref{condLY} is replaced by
\begin{equation}\label{condLY2}
\begin{split}
  & t_1(1-x_1)= W \dfrac{2\pi\hbar^2}{m} B_s[0],\\
  & t_{3'}(1-x_{3'})=(1- W)\dfrac{36\pi\hbar^2}{10m} \fac^{2/3}  B_s[0] .
\end{split}
\end{equation}
This leads to
\begin{widetext}
\begin{equation} \label{ELYO2}
\begin{split}
   \E_c = \E^\mathrm{Sk}_c - \biggl[X_0 a_s[\rho] &+ X_3 \rho^{\alpha} a_s^2[\rho] + 
    (1-W)X_{3'}\rho^{\alpha'} B_s[\rho] \biggr] 
    \biggl[\dfrac{1}{2}\rho^2 - \sum_{q=n,p}\rho_q^2 \biggr] \\
    & - W X_1 B_s[\rho] 
    \biggl[\dfrac{1}{2}\rho\tau+\dfrac{3}{8}(\vec{\nabla}\rho)^2 - \dfrac{1}{4}\vec{J}^{\,2} - \sum_{q=n,p}\Bigl(\rho_q\tau_q+\dfrac{3}{4} (\vec{\nabla}\rho_q)^2\Bigr) \biggr],
\end{split}
\end{equation}
\end{widetext}
where
\begin{equation}
   X_{3'}=\dfrac{3\pi\hbar^2}{5m}\fac^{2/3}.
\end{equation}
$\E^\mathrm{Sk}_c$ now includes a second density--dependent term $\E_{3'}$ with $x_{3'}=1$.

\begin{figure}[htbp]
\begin{center}
\includegraphics[width=\columnwidth]{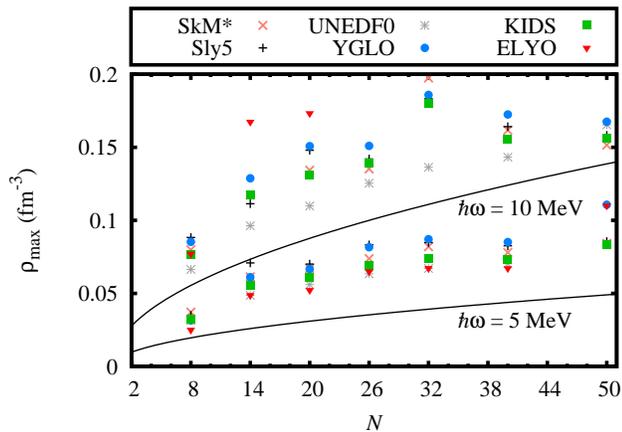}
\caption{Maximal density at the Thomas-Fermi approximation as a function of the neutron number for the two trap frequencies $\hbar \omega=5$ MeV (bottom solid line) and $10$ MeV (top solid line). We also show for the two cases, systematically, the maximal densities obtained for the different considered functionals. Note that the EDF results are always above the Thomas-Fermi value due to the attractive self--consistent mean field.}
\label{fig:densmax}
\end{center}
\end{figure}

The new density--dependent term $\E_{3'}$ does not affect the EOS for PNM as, by construction, it recombines with $\E_1$ so that the original condition Eq. \eqref{condLY} to match the Lee-Yang expansion is recovered for any value of $W$. On the other hand, the undetermined parameter $t_{3'}$ enters in the SNM EOS. By analogy to the YGLO and KIDS cases, we therefore impose the SNM EOS to also remain unchanged by the introduction of this new term, which gives
\begin{equation}
\begin{split}
  & t_1 = W t_1^0, \\
  & t_{3'}= (1-W)\dfrac{9}{5} \left(\dfrac{3\pi^2}{2}\right)^{2/3} t_1^0,
\end{split}
\end{equation}
where $t_1^0$ denotes the value of $t_1$ resulting from the fit of SNM EOS obtained for $W=1$ in Ref. \cite{ELYO}. The values of the other parameters, $t_0$ and $t_3$, are taken to be those of the initial version of the ELYO EDF.  This dispenses a supplementary adjustment, but is possible only because of the pure $s$--wave character ($t_2=0$) of the functional.
%
%
%
\section{Results and discussions} \label{SecRes}
%
%
The above--described fitting procedure for the three functionals yields the splitting coefficient $W$, the spin--orbit coupling constant $V_\mathrm{so}$, and the pairing strength $V_\mathrm{pp}$, reported in Table \ref{tab_CCfit} together with those corresponding to some Skyrme EDFs. 

\begin{table}[b]\begin{tabular*}{\linewidth}{@{\extracolsep{\fill}}crrrrrr}
\hline
\hline
                 & YGLO      &   KIDS     &    ELYO   &    Sly5   & SkM*    & UNEDF0  \\
\hline
$W$              & $-0.084$  & $ 0.110$   & $ 0.396$  &     --    &   --    &   --    \\     
$V_\mathrm{so}$  & $ 138.2$  & $ 110.0$   & $  55.0$  & $  125.0$ & $130.0$ & $ 91.3 $  \\
$V_\mathrm{pp}$  & $-275.1$  & $-183.9$   & $-152.5$  & $ -213.1$ & $-233.9$& $ -170.4$    \\

\hline
\hline
\end{tabular*}
\caption{Splitting parameter, spin--orbit coupling constant (in MeV fm$^5$), and pairing strength (in MeV fm$^3$) ensuing from the fit described in Sect. \ref{EDFforDrops}.  For the KIDS EDF, $V_\mathrm{so}$ results from an adjustment on the binding energies of ${}^{48}$Ca and ${}^{208}$Pb \cite{KIDS5}. Also shown are values corresponding to some Skyrme functionals for comparison.}\label{tab_CCfit}
\end{table}

Table \ref{tab_CCfit} shows that, whereas the KIDS functional admits rather standard $V_\mathrm{so}$ and $V_\mathrm{pp}$ values, the YGLO functional leads to a slightly higher pairing parameter and the ELYO functional has a lower spin--orbit strength. Nonetheless, these differences may be consequences of the fact that the parameters of YGLO and ELYO stem from an adjustment on neutron drops properties, and not on nuclei. Knowing whether these values are features of the functionals themselves or consequences of the fitting procedure would require us to consider nuclei, which is out of the scope of the present study.

Before describing the obtained results, we estimate the density 
range explored with two trap frequencies under consideration. For this, we compute the maximal density obtained for the trapped free Fermi gas (at the Thomas-Fermi approximation \cite{Lip08}) and we plot it in Fig. 
\ref{fig:densmax}
as a function of the neutron number for the two trap frequencies $\hw=5$ and $10$ MeV. Such density values are compared to those corresponding to the interacting gas of neutrons (described with the adjusted functionals). 
Note that the maximal density for the system of interacting neutrons is always higher due to the mutual attraction between neutrons (mean field).

For $\hw=5$ MeV (lower compression), all functionals globally lead to the same maximal density.
However, more important differences are observed for larger compression, $\hw=10$ MeV.  These differences can partially be understood from 
the corresponding EOSs for PNM (Fig. \ref{fig_EoSMatter}). This is, for instance, the case for the ELYO functional that deviates significantly 
from other EOSs at densities higher than 0.1 fm$^{-3}$: the corresponding maximum density becomes much more important in this density region where, correspondingly, the ELYO PNM EOS predicts a much more bound system.  

\begin{figure}[htbp]
\begin{center}
\includegraphics[width=0.9\columnwidth]{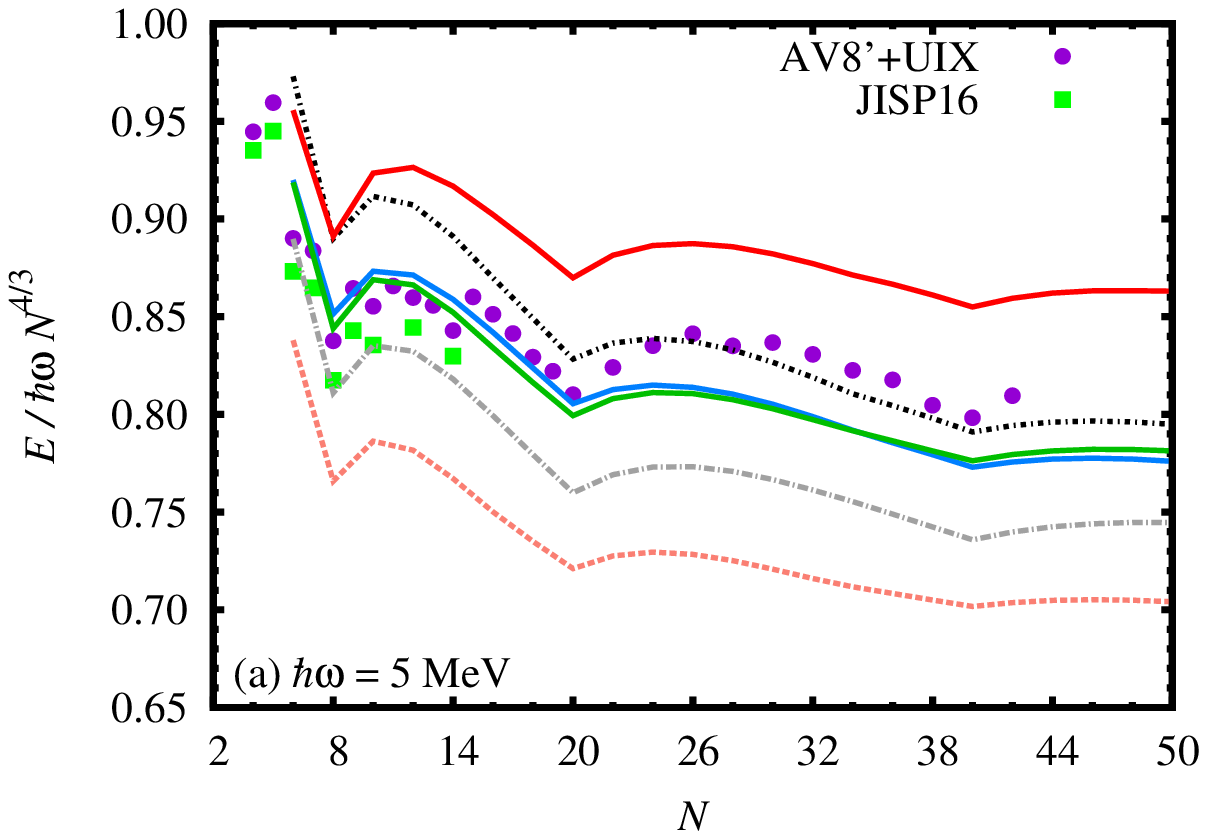}
\includegraphics[width=0.9\columnwidth]{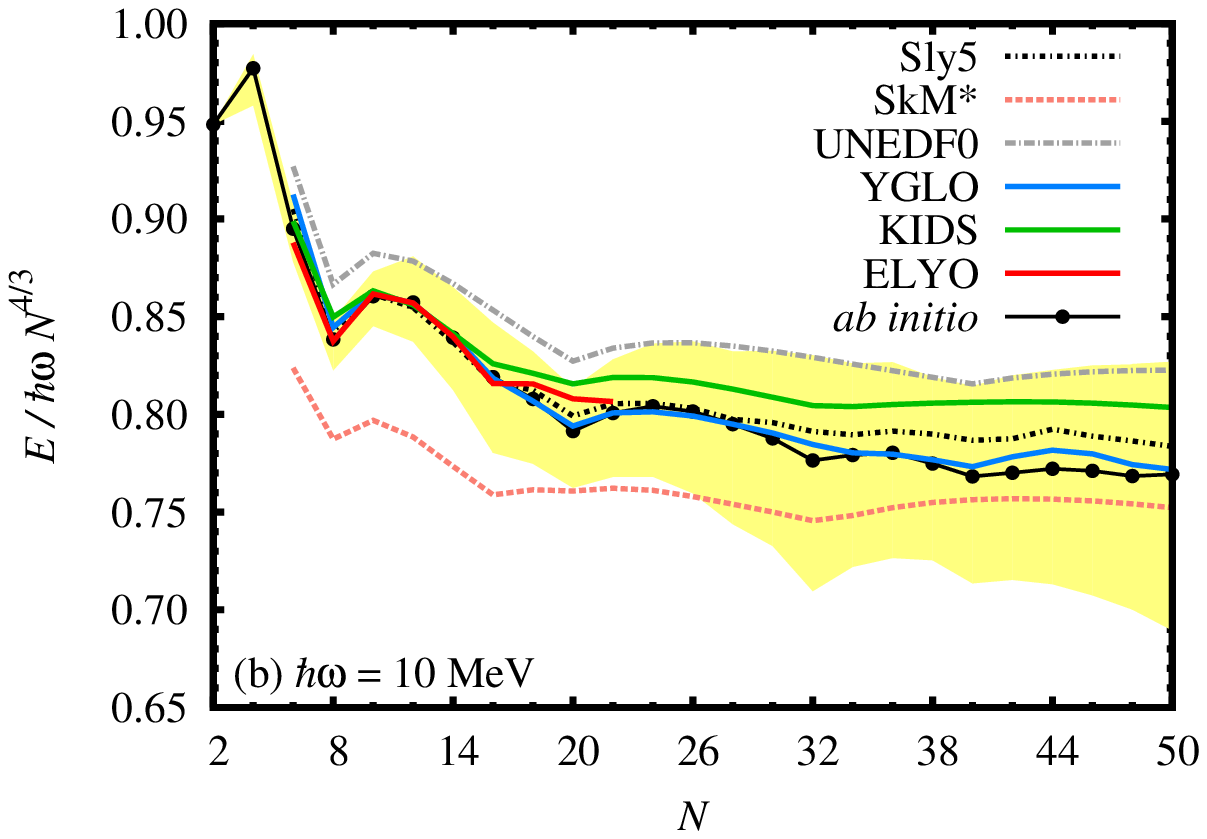}
\caption{Energies of neutron drops as a function of $N$ obtained for (a) $\hw=5$ and (b) 10 MeV with various EDFs compared to \textit{ab initio} results.}
\label{fig_EoSDrop}
\end{center}
\end{figure}

\begin{figure}[htbp]
\begin{center}
\includegraphics[width=0.9\columnwidth]{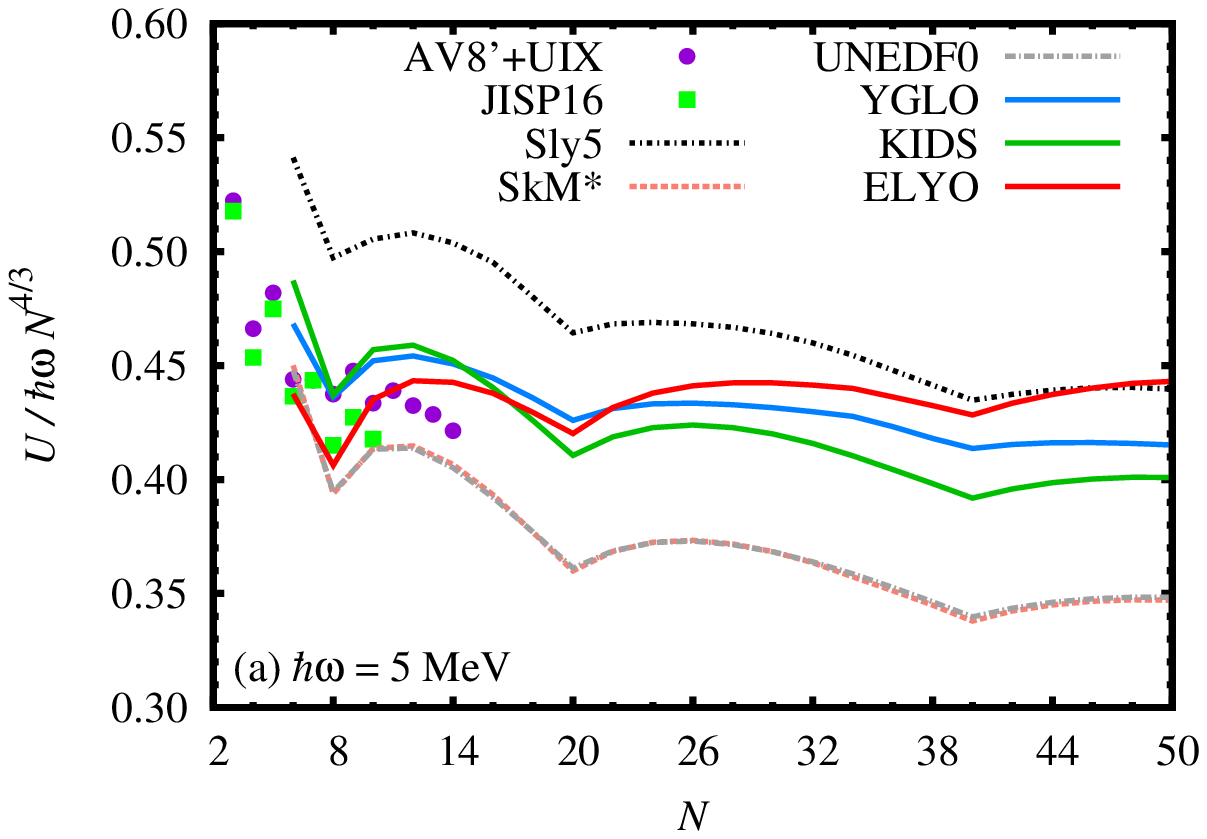}
\includegraphics[width=0.9\columnwidth]{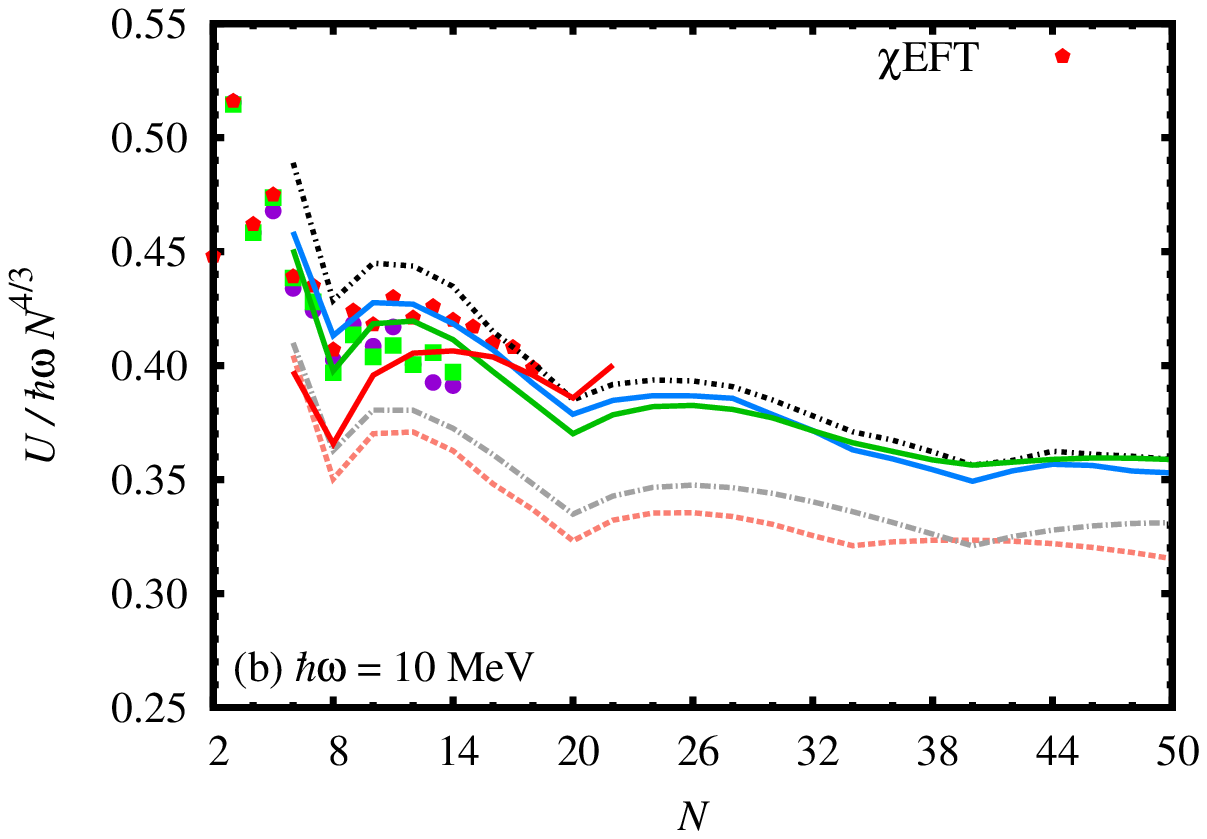}
\caption{Internal energies $U$ of neutron drops for (a) $\hw=5$ and (b) 10 MeV as obtained with various EDFs compared to \textit{ab initio} results.}
\label{fig_UDrop}
\end{center}
\end{figure}

\subsection{Energetic properties and densities}
We now compare the results for various quantities predicted from the KIDS, YGLO, and ELYO EDFs (using the parameters of Table \ref{tab_CCfit}) to those obtained with commonly used Skyrme EDFs and to available \textit{ab initio} calculations. Figure \ref{fig_EoSDrop} displays the evolution of the scaled energies $E/\hbar\omega N^{1/3}$ as a function of the neutron number $N$ for $\hw =5$ [Ref. \ref{fig_EoSDrop}(a)] and $10$ [Ref. \ref{fig_EoSDrop}(b)] MeV. 
For $\hw =10$ MeV, the YGLO and ELYO functionals provide a rather good reproduction of the {\it ab initio} reference points for $N<20$. This is not surprising since both have been explicitly adjusted to reproduce this region of particle number. For the KIDS case, where only the pairing strength was adjusted, and for the Sly5 case, where no adjustment was done, results are also consistent with the {\it ab initio} ones. Note that the KIDS functional gives slightly higher energies as $N$ increases.  
Unexpectedly, the SkM* and UNEDF0 parametrizations, which provide a rather similar description of PNM and SMN EOSs (Fig. \ref{fig_EoSMatter}), exhibit marked discrepancies with respect to one another but also with respect to other functionals and to the {\it ab initio} reference curve. One underbinds and the other one overbinds systematically the droplets of neutrons. 
\begin{figure*}[t]
\begin{center}
\includegraphics[width=0.8\columnwidth]{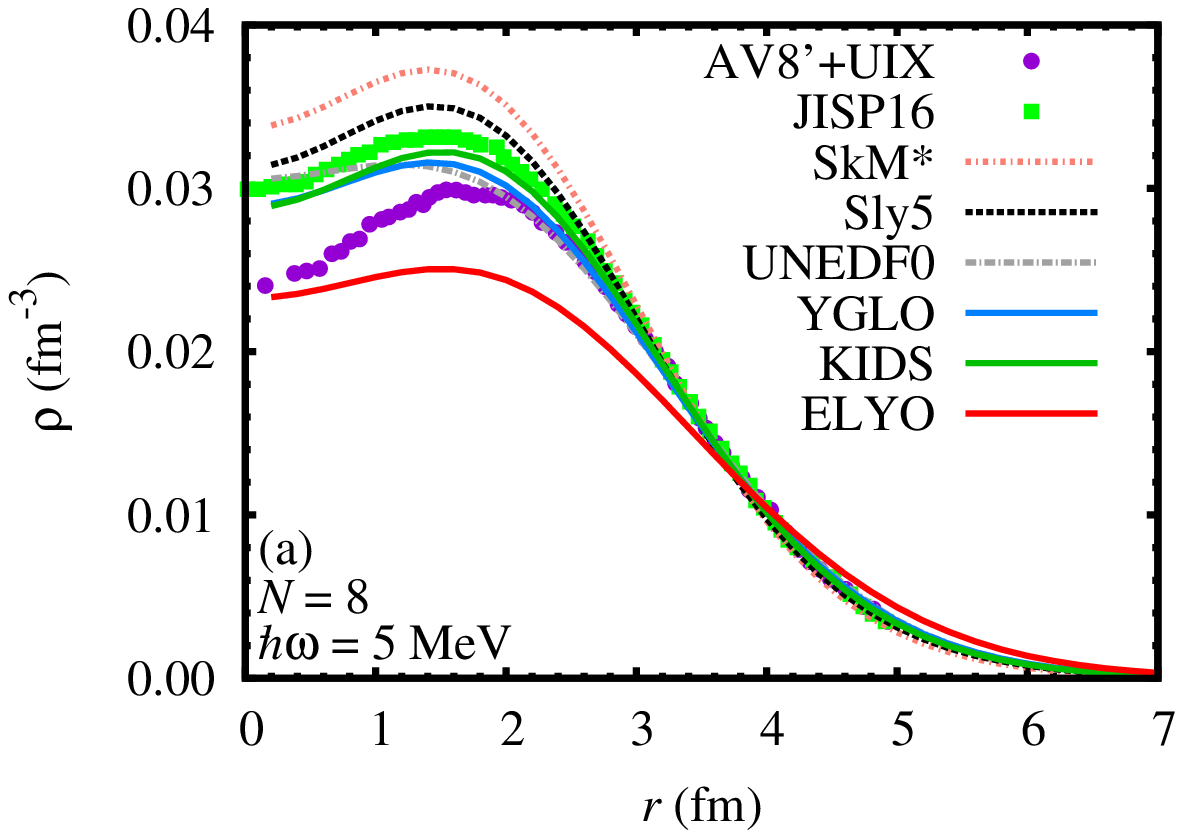}\includegraphics[width=0.8\columnwidth]{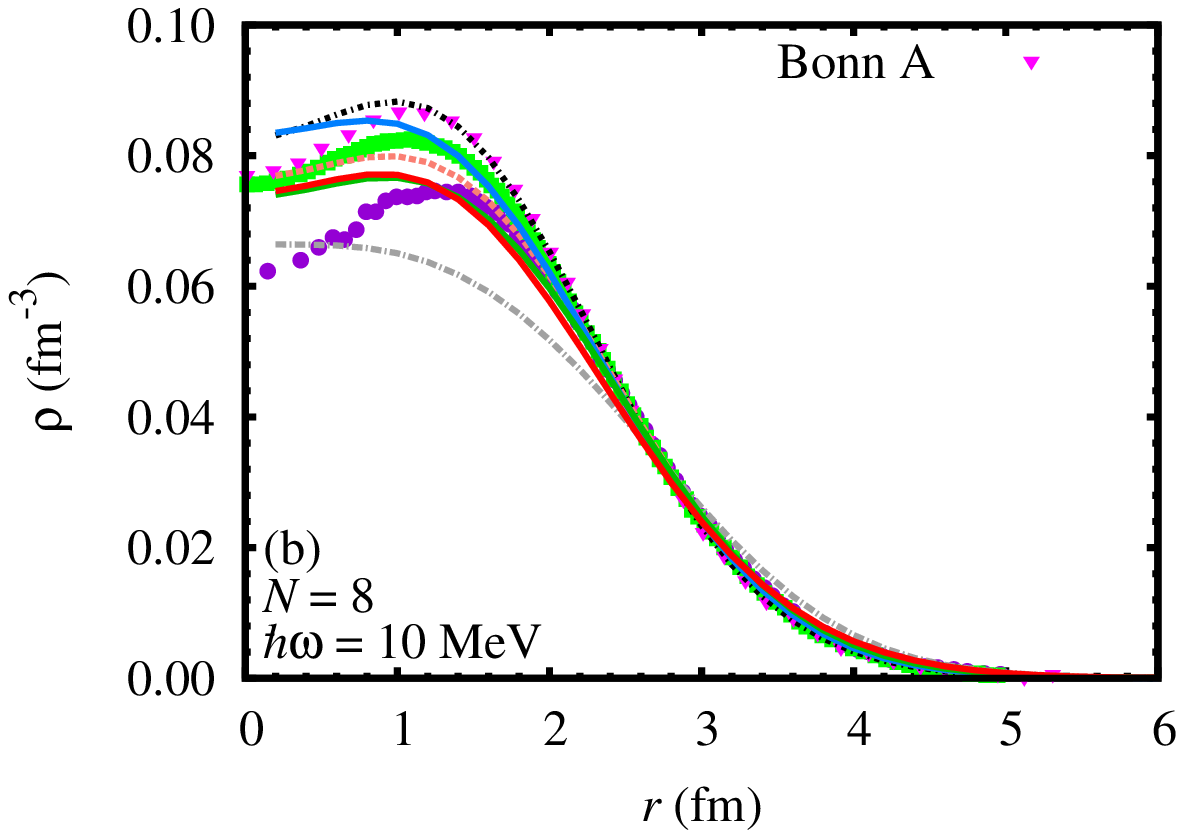}\\ 
\includegraphics[width=0.8\columnwidth]{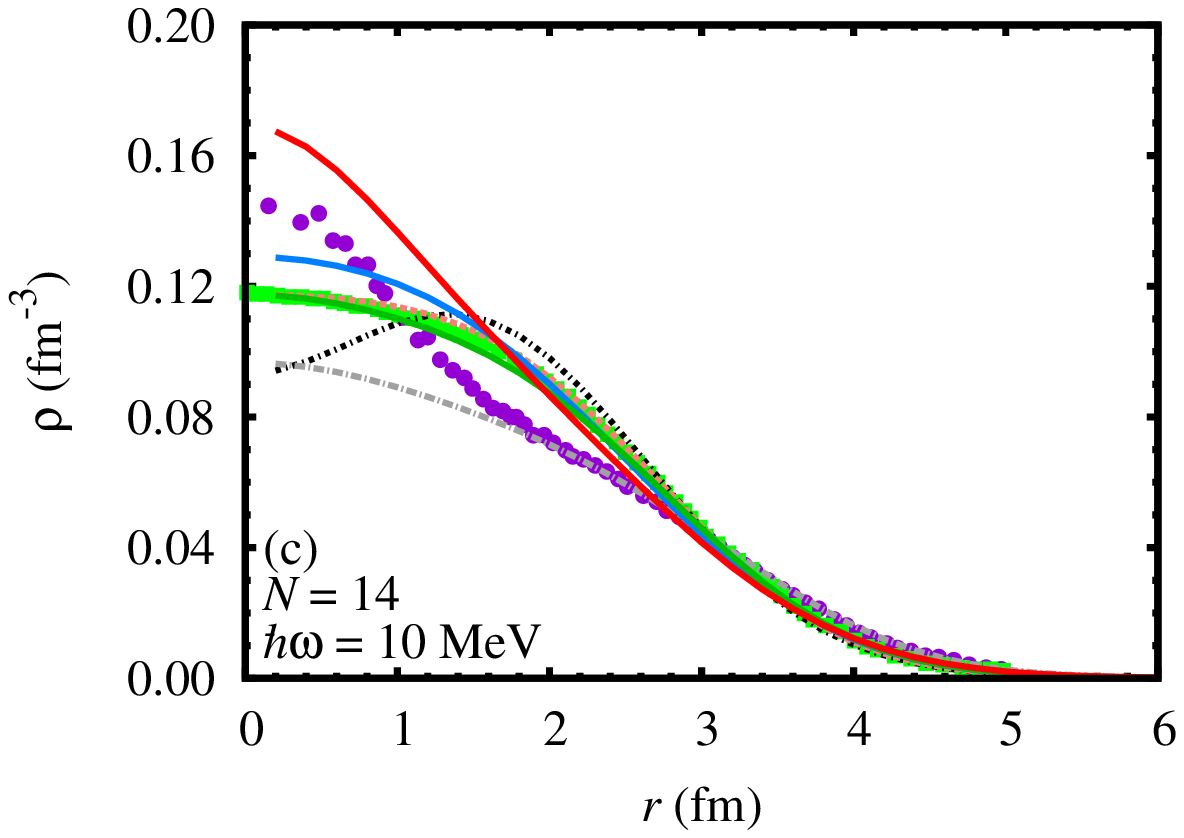}\includegraphics[width=0.8\columnwidth]{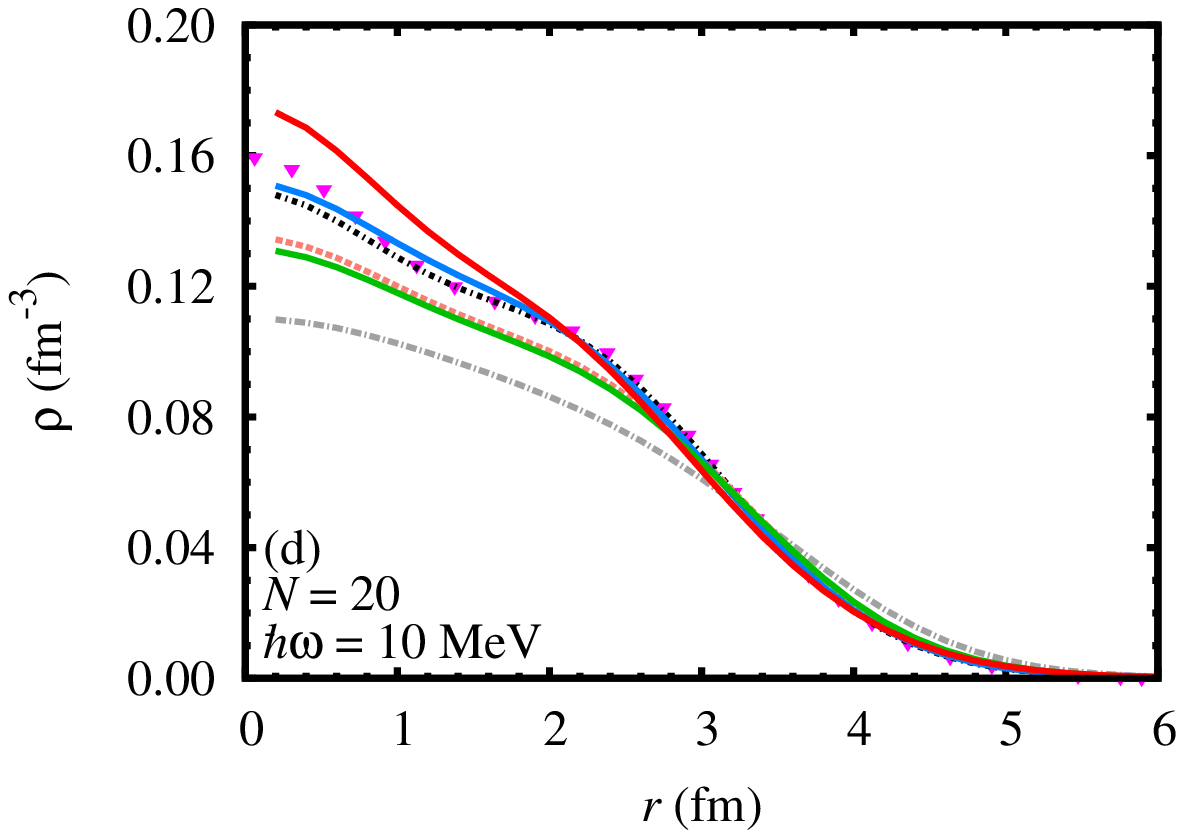}\\
\includegraphics[width=0.8\columnwidth]{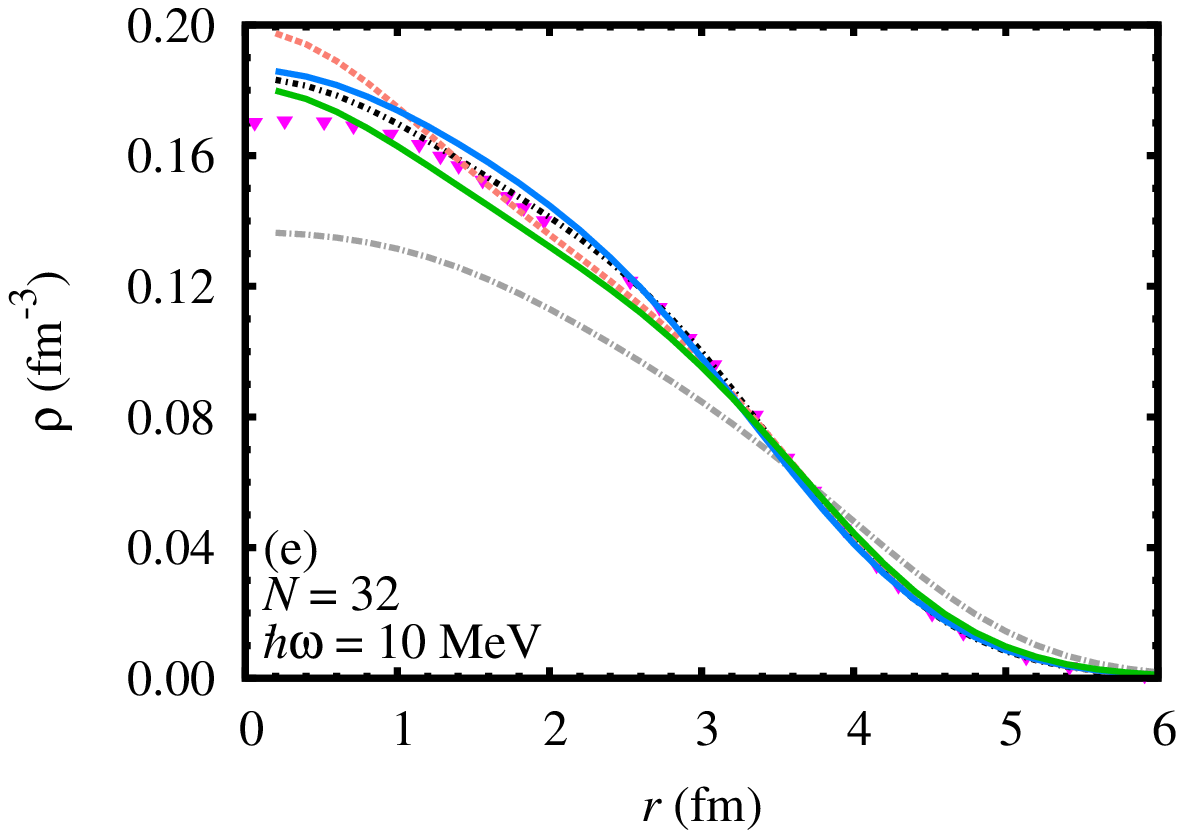}\includegraphics[width=0.8\columnwidth]{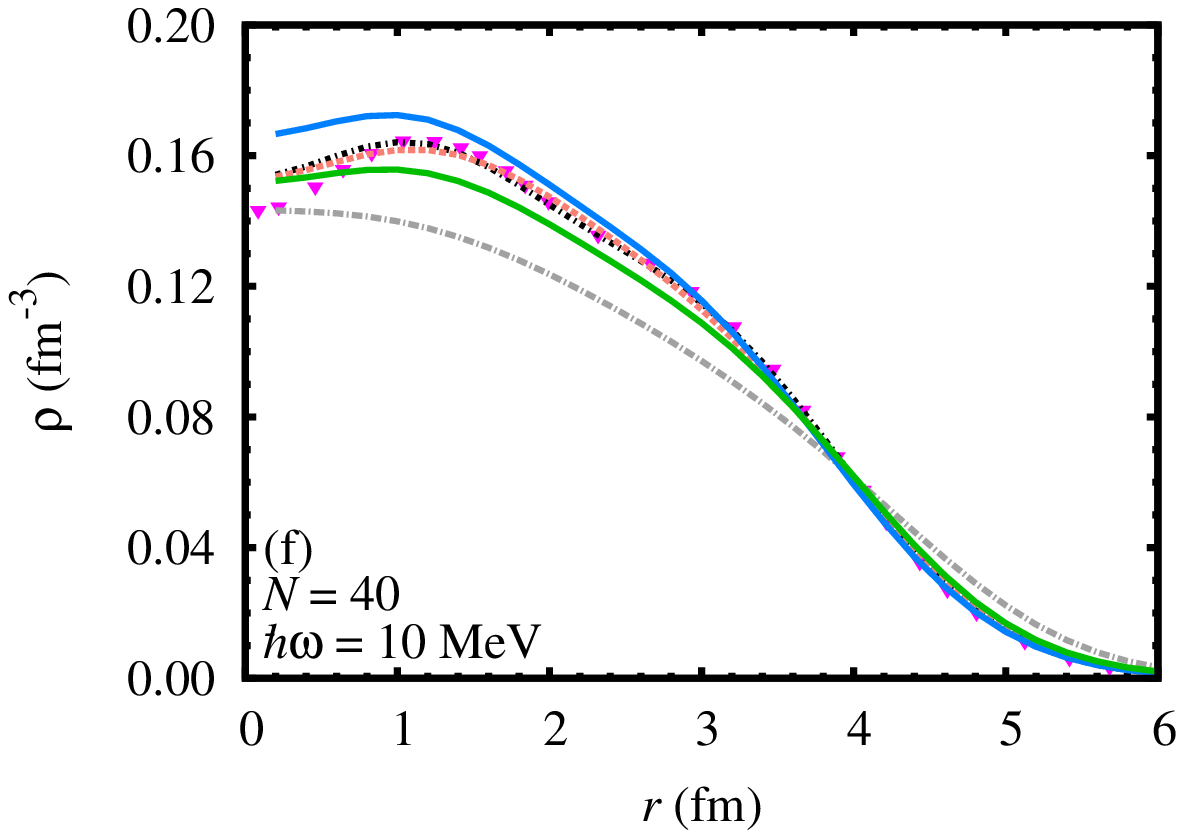}\\
\caption{Density profiles as a function of the distance from the center of the trap obtained from the different functionals 
for (a) ($N=8, \hw =5$ MeV), (b) ($N=8, \hw =10$ MeV), (c) ($N=14, \hw =10$ MeV), (d) ($N=20, \hw =10$ MeV), (e) ($N=32, \hw =10$ MeV), and (f) ($N=40, \hw =10$  MeV). The {\it ab initio} results are extracted from Ref. \cite{NCSMQMC} (purple circles and green squares), and Ref. \cite{tensor2} (pink triangles).}\label{fig_DensDrop}
\end{center}
\end{figure*}
For the ELYO functional, while satisfactory energies are obtained up to $N=22$, the treatment of heavier systems cannot be carried out owing to numerical instabilities in the solution of the HFB equations.  This behavior turns out to be consistent with the ELYO EOS for neutron matter. Indeed, evaluating the number of neutrons per unit volume with the calculated radii, we find that the range $6\leq N\leq 20$ corresponds to densities between 0.07 and 0.16 fm${}^{-3}$ where the three EFT--inspired EDFs provide similar energies per particle. The limit $N=22$, from which instabilities arise, corresponds to  $\rho > 0.2$ fm${}^{-3}$ where the ELYO EOS strongly departs from the others.
This is consistent with Fig. \ref{fig:densmax} where we show the maximal densities $\rho_\mathrm{max}$ associated to the functionals: The abrupt jump in the value of $\rho_\mathrm{max}$ for ELYO from 0.174 fm${}^{-3}$ at $N=20$ to 0.295 fm${}^{-3}$ (not shown in the figure) at $N=22$ may be viewed as a warning sign of the instabilities beyond $N=22$.

For the weaker trapping potential, Fig.  \ref{fig_EoSDrop}(a), large differences between different EDFs are observed for both small and large neutron numbers. This is particularly evident for the YGLO, Sly5, KIDS, and ELYO cases that were almost superposed for $N<14$ in Fig. \ref{fig_EoSDrop}(b). The KIDS  and YGLO results are still rather close to one another and not far from the set of {\it ab initio} results shown for $\hw=5$ MeV. In this case, the Sly5 and ELYO EDFs overestimate the energy for small particle numbers. As $N$ increases, this feature persists for ELYO whereas the discrepancy with {\it ab initio} results diminishes for the Sly5 case.  
The SkM* EDF, as for the higher trapping frequency, always leads to underbound drops. 
We observe that, for $\hbar \omega = 5$ MeV, ELYO results are obtained for the whole window of neutron numbers. It turns out that, for $\hbar \omega = 5$ MeV, the range $8\leq N\leq 50$ is equivalent to $0.03 \leq \rho \leq 0.1$ fm$^{-1}$. In this density window the ELYO EOS for PNM is not strongly different from the other EOSs, and still provides reasonable results. In this density region, the ELYO PNM EOS is located at higher energies compared to YGLO and KIDS. The same behavior is observed for the droplet energies (Fig. \ref{fig_EoSDrop}).

Figure \ref{fig_UDrop} presents the evolution of the internal energy 
\begin{eqnarray}
U=\displaystyle E-\int\d^3\vec{r}\E_\mathrm{\omega}(\vec{r}), \label{eq:u}
\end{eqnarray}
that is the total energy minus the contribution of the external trap, with the number of neutrons for the two trap frequencies 5 and 10 MeV. 
We note that the comparison with respect to {\it{ab initio}} calculations may lead in this case to different conclusions for each of the functional, compared to what is found in Fig. \ref{fig_EoSDrop}. 
The difference between the energy $E$ and the internal energy $U$ is essentially coming from the density profile involved in $\E_\mathrm{\omega}(\vec{r})$ (see Eq. \eqref{edffull}). Such a change in the trend of the results for the internal energy, compared to the total energy, must thus be produced by different density profiles obtained for each functional.  
Figure \ref{fig_DensDrop} displays the obtained density profiles. 
We remark that density profiles deduced from {\it ab initio} calculations might have significant discrepancies from one another, which makes any comparison only qualitative.

\subsection{Mean fields and pairing gaps}
\begin{figure*}[htbp]
\begin{center}
\includegraphics[width=0.8\columnwidth]{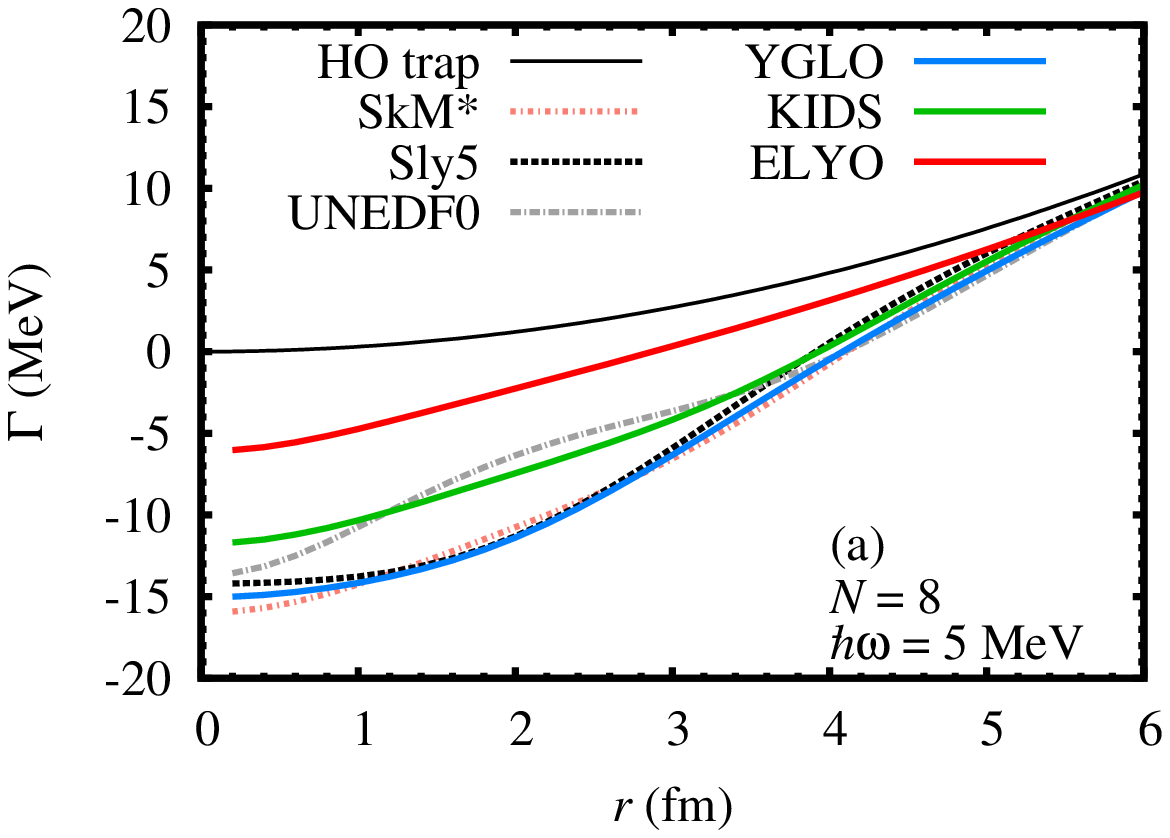}\includegraphics[width=0.8\columnwidth]{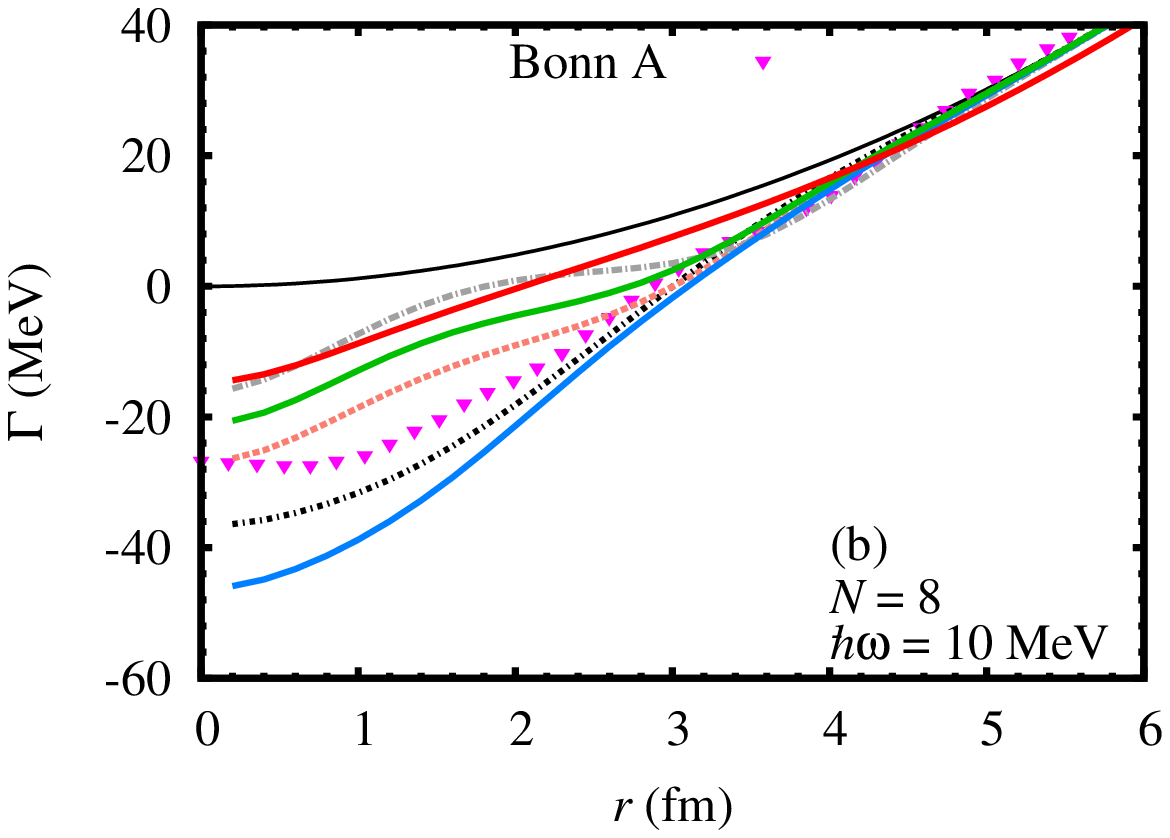}\\
\includegraphics[width=0.8\columnwidth]{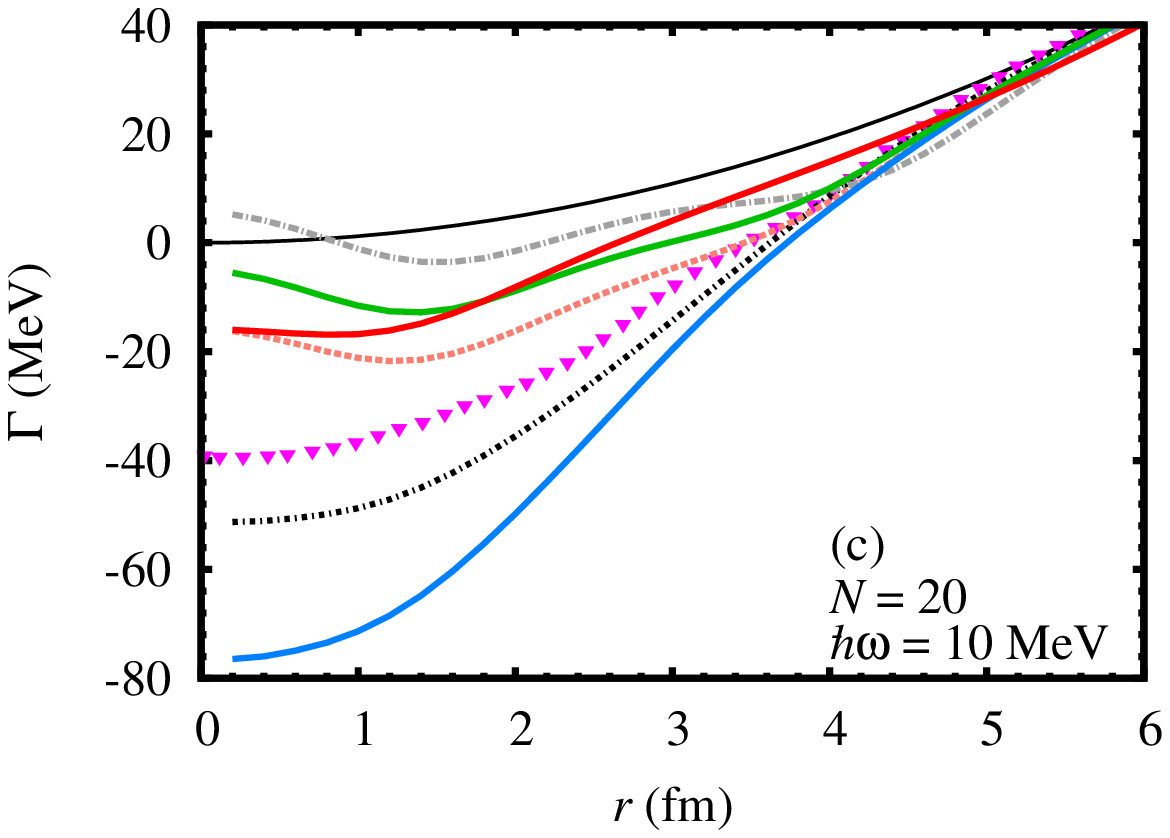}\includegraphics[width=0.8\columnwidth]{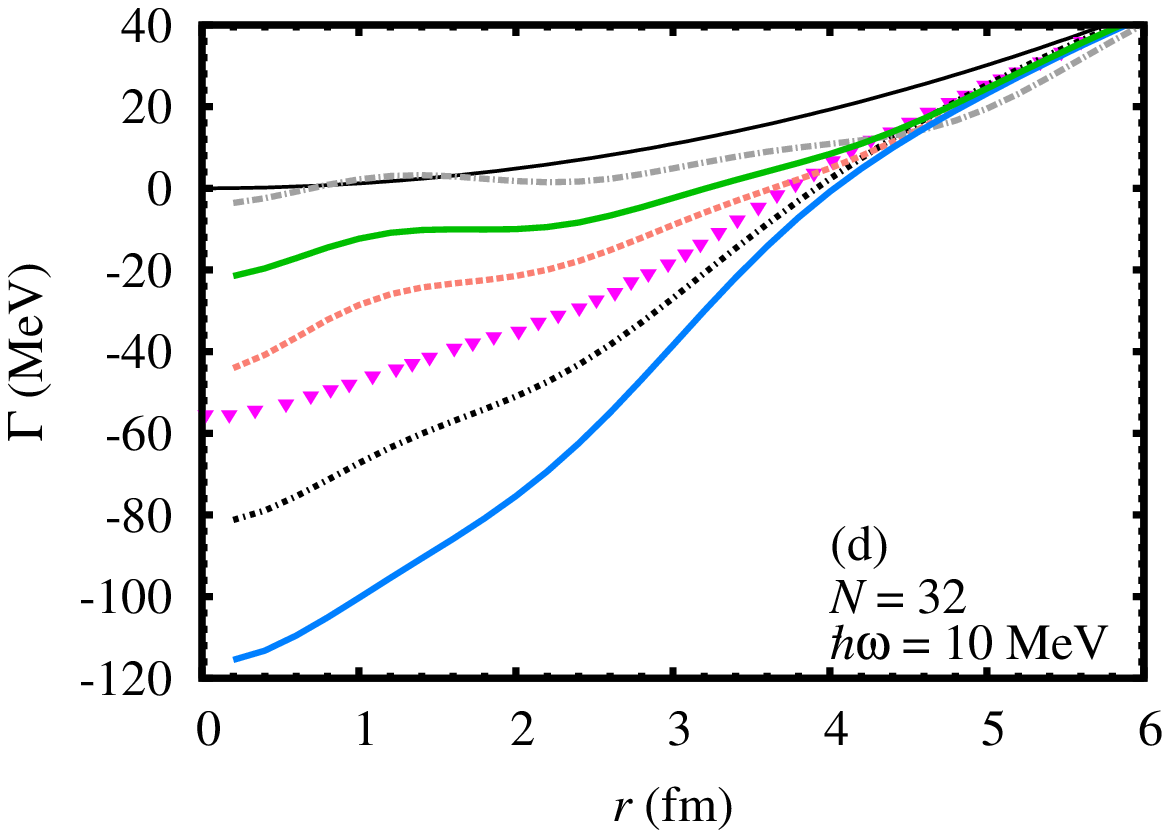}\\
\includegraphics[width=0.8\columnwidth]{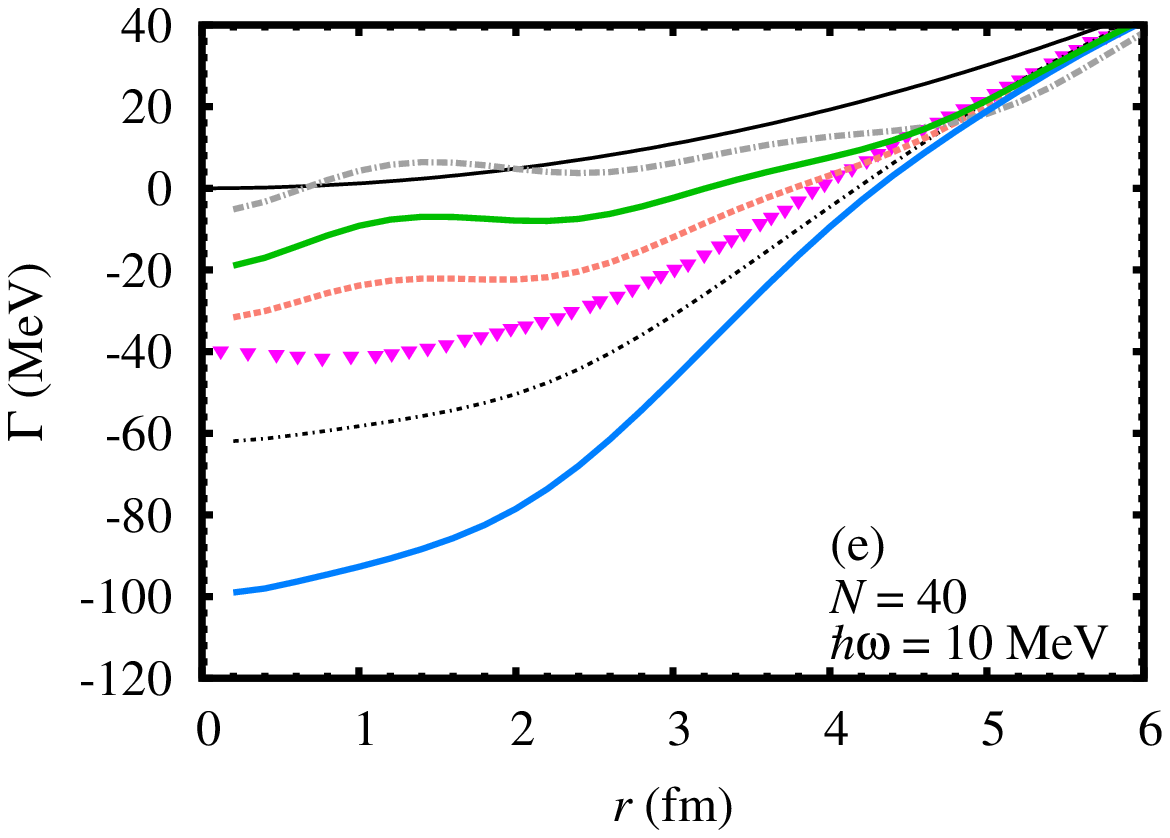}\includegraphics[width=0.8\columnwidth]{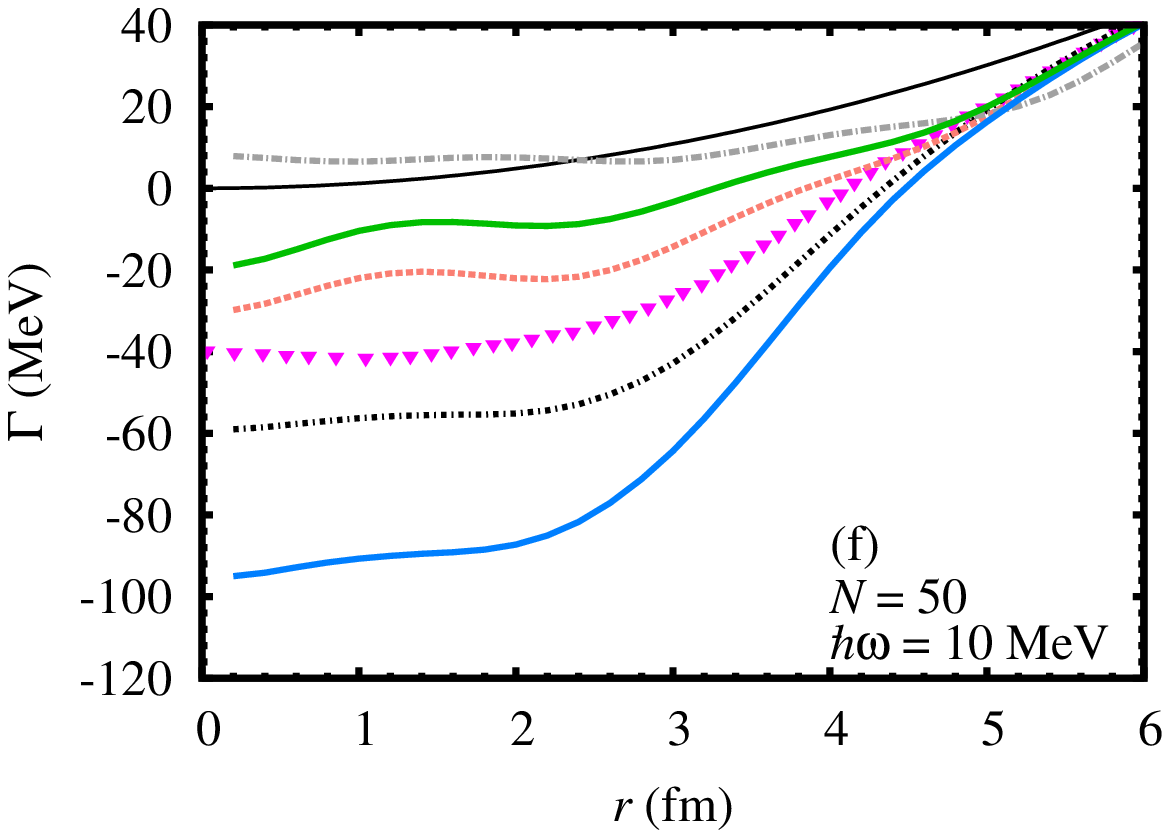}
\caption{Hartree-Fock potentials from the different functionals for (a) ($N=8, \hw =5$ MeV), (b) ($N=8, \hw =10$ MeV), (c) ($N=20, \hw =10$ MeV), (d) ($N=32, \hw =10$ MeV), (e) ($N=40, \hw =10$ MeV), and (f) ($N=50, \hw =10$ MeV).
RHFB results based on the Bonn A interaction (pink triangles) are from Ref. \cite{tensor2}. Note that for (e) and (f), the ELYO functional 
do not lead to converged results and therefore, it is not shown.}
\label{fig_FieldDrop}
\end{center}
\end{figure*}

\begin{figure}[t]
\begin{center}
\includegraphics[width=0.9\columnwidth]{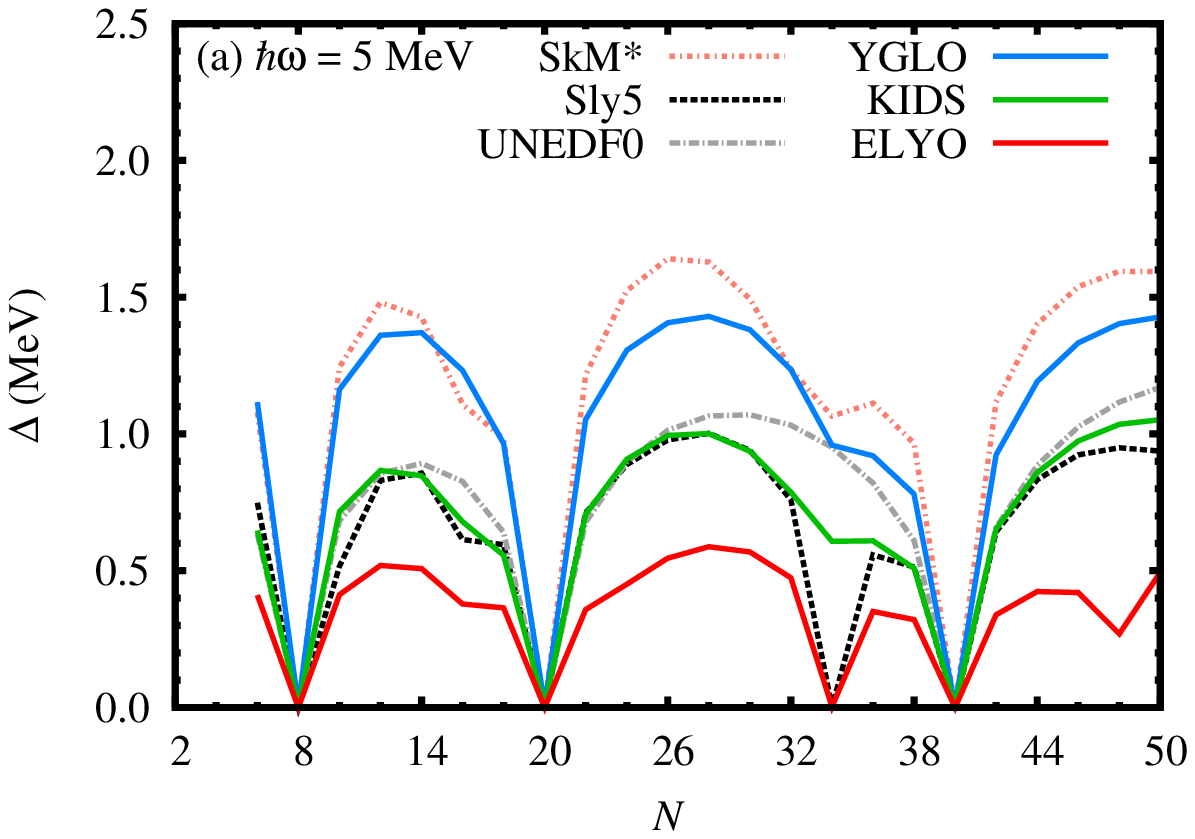}
\includegraphics[width=0.9\columnwidth]{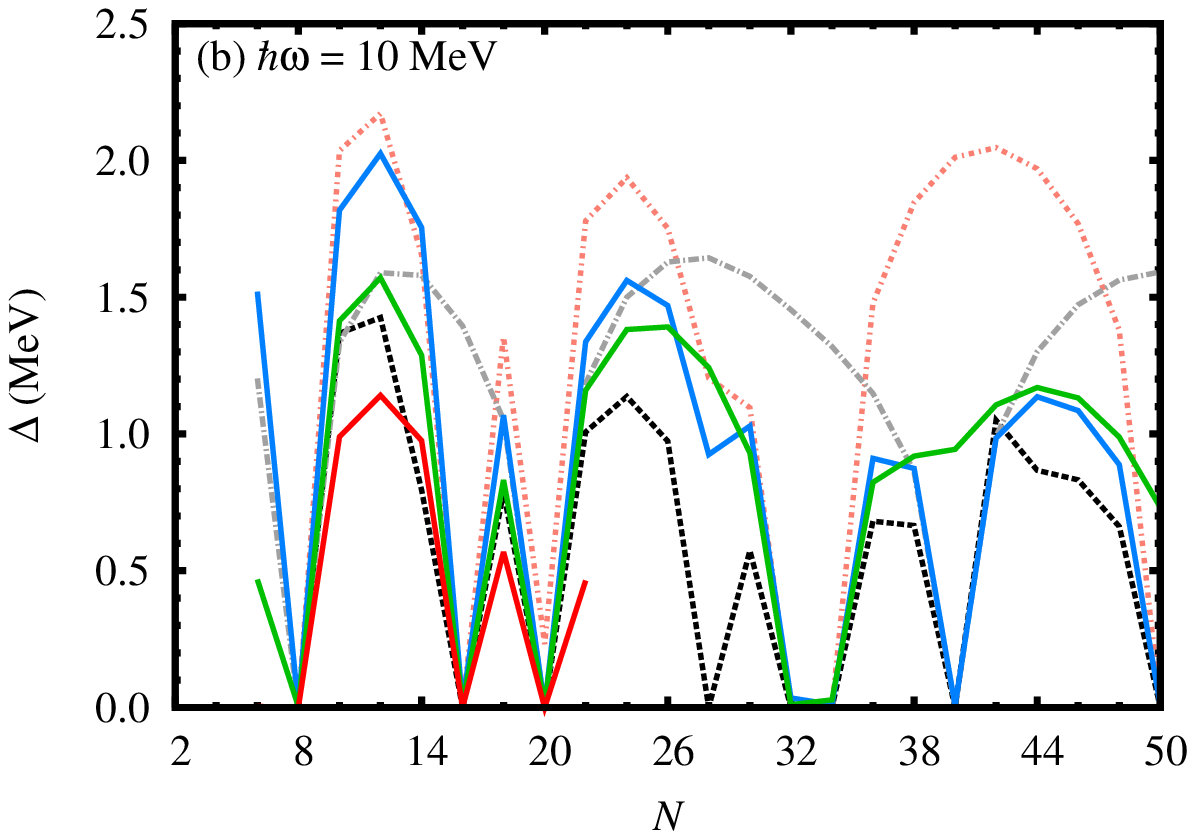}
\caption{Mean pairing gaps of neutron drops for (a) $\hw=5$ and (b) 10 MeV obtained with the EDFs  considered in this work.}
\label{fig_DDrop}
\end{center}
\end{figure}

One of the advantages of the EDF theory compared, for instance, to {\it ab initio} many--body methods is that it gives direct access to quasiparticle properties such as the one--body self--consistent mean field or the pairing gap. These quantities are, respectively, reported in Figs.  \ref{fig_FieldDrop} and \ref{fig_DDrop} for the two frequencies of the trap and for various particle numbers.  Results obtained using the relativistic Brueckner-HF calculations with the Bonn A interaction are also shown for comparison when available \cite{tensor,tensor2}. 

Focusing first on the mean field, we observe a very large dispersion of the results depending on the functional, even for cases that lead to similar PNM EOSs.  These differences can be partially attributed to the different sharing of the energy between volume and surface terms. 

If we now consider the pairing gaps and compare the three EFT--inspired functionals, we notice that, for the trap of frequency $\hw=5$  MeV, the pairing gap is the largest for the YGLO EDF and the weakest for the ELYO one, the KIDS case being intermediate. This reflects the ordering of the corresponding pairing strengths (Table \ref{tab_CCfit}). On the other side, for the  10--MeV trap, such a behavior occurs only for the lightest systems. Starting from  $N\sim20$, YGLO and KIDS provide comparable values for the pairing gaps. As the YGLO pairing strength is more important, this can be explained only by a larger energy--level  spacing in the YGLO case. Indeed, one may see in Fig. \ref{fig_FieldDrop} that, for this trap frequency, passing from $N=8$ to $N=20$, the depth of the KIDS potential remains more or less the same (20 MeV) whereas that of the YGLO functional is strongly enhanced (from 50 to 80 MeV). Unfortunately, benchmark data are missing for the pairing gaps and we cannot compare our results with microscopic values.

\subsection{Effective mass}

The effective mass for neutron matter is poorly constrained in the EDF approach (see Fig. 6 of Ref. \cite{lacroix2}). 
Let us mention a recent work where, for the KIDS case, a procedure to extract a functional for nuclei from a functional tailored to provide a given EOS is discussed employing the effective mass \cite{KIDS4}. 

\begin{figure}[t]
\begin{center}
\includegraphics[width=0.9\columnwidth]{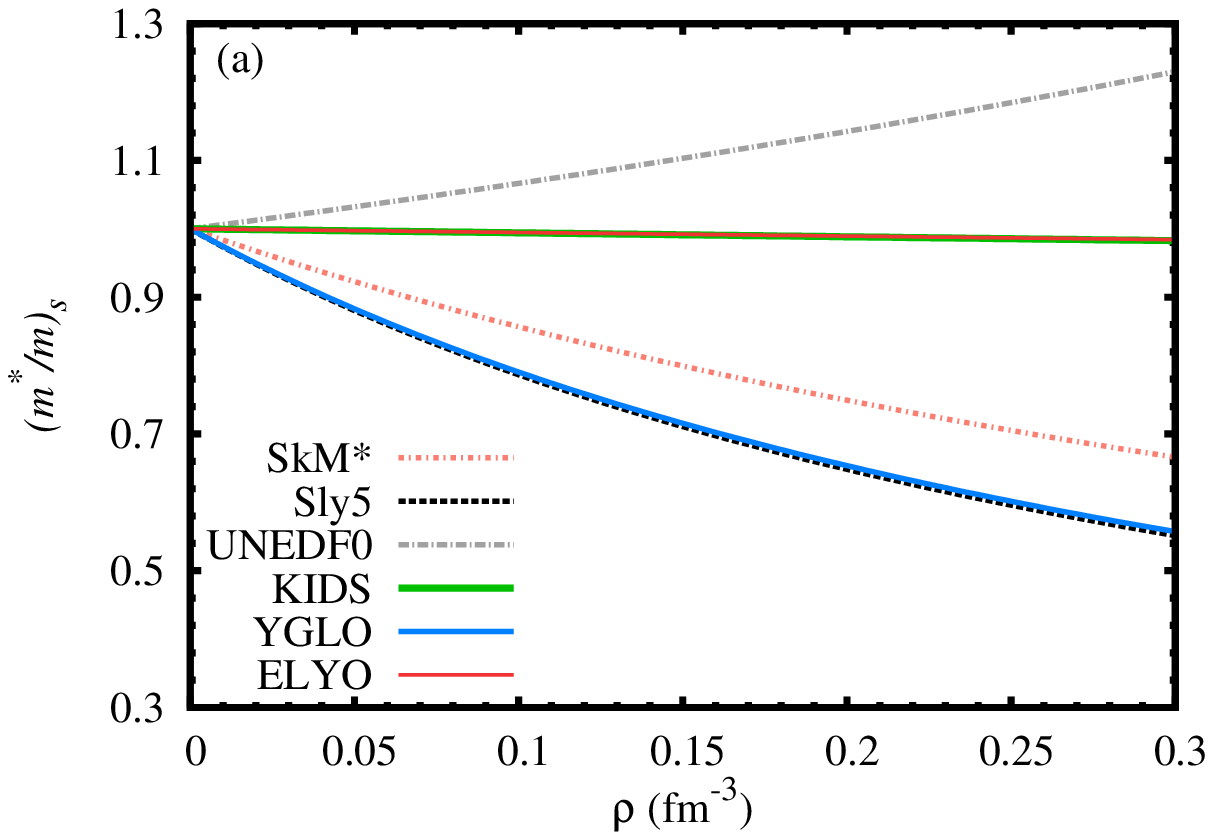}
\includegraphics[width=0.9\columnwidth]{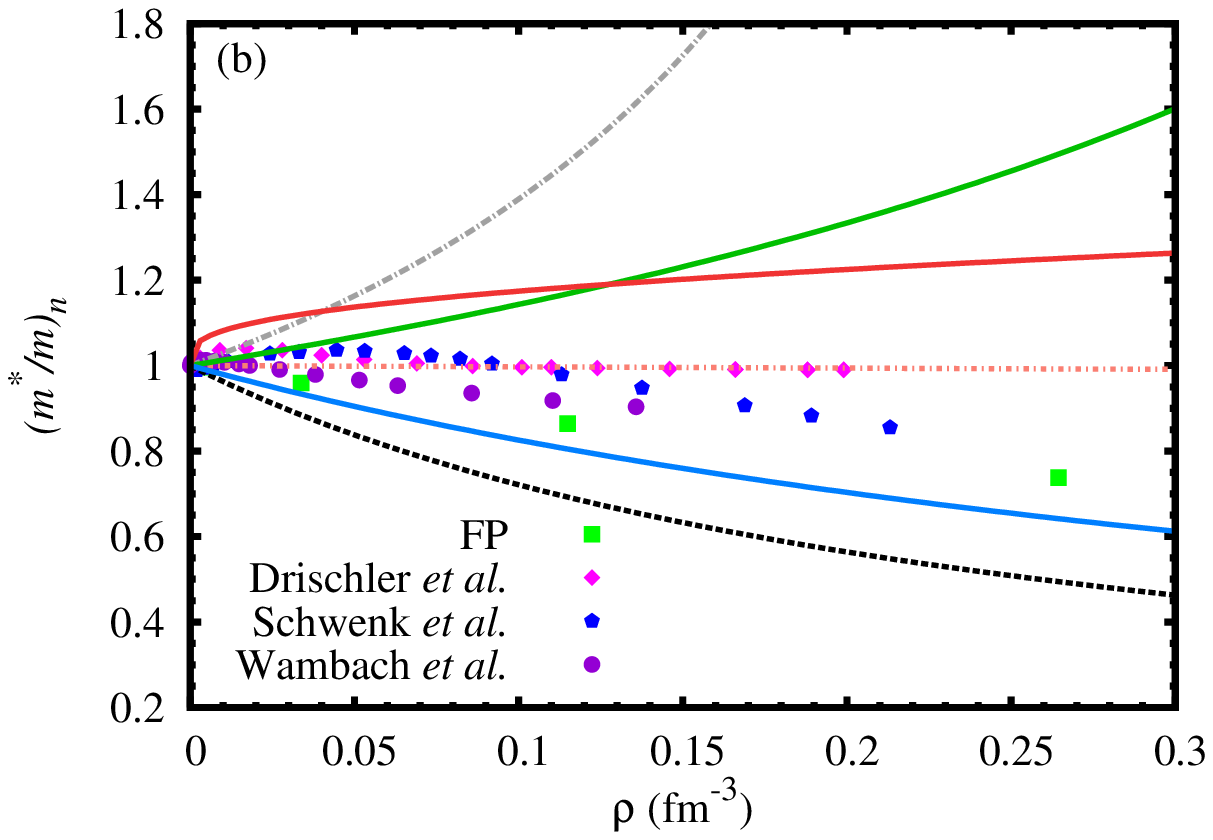}
\caption{(a) Isoscalar (SNM) and (b) neutron (PNM) effective masses as a function of the density obtained with various EDFs (lines) compared to \textit{ab initio} values extracted from FP (Ref. \cite{QMCFP}, green squares), Drischler \textit{et al}. (Ref. \cite{DSS}, pink diamonds), Schwenk \textit{et al}. (Ref. \cite{SFB}, blue pentagons), and Wambach \textit{et al}. (Ref. \cite{WAP}, purple circles). The YGLO and KIDS curves ensue from the optimal value of $W$ shown in  Table \ref{tab_CCfit}. Note that in panel (a), the KIDS and ELYO effective masses are identical.}
\label{fig_mstar}
\end{center}
\end{figure}

The evolution of the isoscalar (SNM) and the neutron (PNM) effective masses with the density (Eq. \eqref{effMass}) for the three functionals discussed here is plotted in Fig. \ref{fig_mstar} and compared to those from the SkM*, Sly5, UNEDF0 parametrizations. Large dispersions are noticed for the standard Skyrme EDFs.
Also shown are \textit{ab initio} estimates for neutron matter from Refs. \cite{QMCFP,DSS,SFB,WAP} that seem, globally, to predict values closer to the bare mass.

In the case of the three EFT--based functionals, the effective mass is strongly impacted by the value of the $W$ parameter stemming from the fitting protocol. Indeed, no splitting ($W=0$) implies that $(m^*/m)_{s,n}$ are constant and equal to 1 at any density in the KIDS and YGLO cases.

With the optimal $W$ coefficients of Table \ref{tab_CCfit}, the YGLO and KIDS neutron effective masses depart from unity in opposite directions but both stay qualitatively close to the \textit{ab initio} range of values up to $\rho=0.02$ fm${}^{-3}$. 
For the ELYO functional, the splitting ($W$) entails a significant reduction of $(m^*/m)_{n}$ that becomes smaller that the 
KIDS value around the saturation density whereas we observed that it remains largely above when the splitting is absent (which corresponds to $W=1$ for ELYO).

The effect of the splitting on the SNM effective masses is quite different. $W$ does almost not affect $(m^*/m)_{s}$ for the KIDS and ELYO EDFs. In contrast, it modifies the YGLO isoscalar effective mass so that its curve follows the Sly5 one and approaches the commonly admitted value of 0.7 around the saturation density.

\begin{figure}[t]
\begin{center}
\includegraphics[width=0.9\columnwidth]{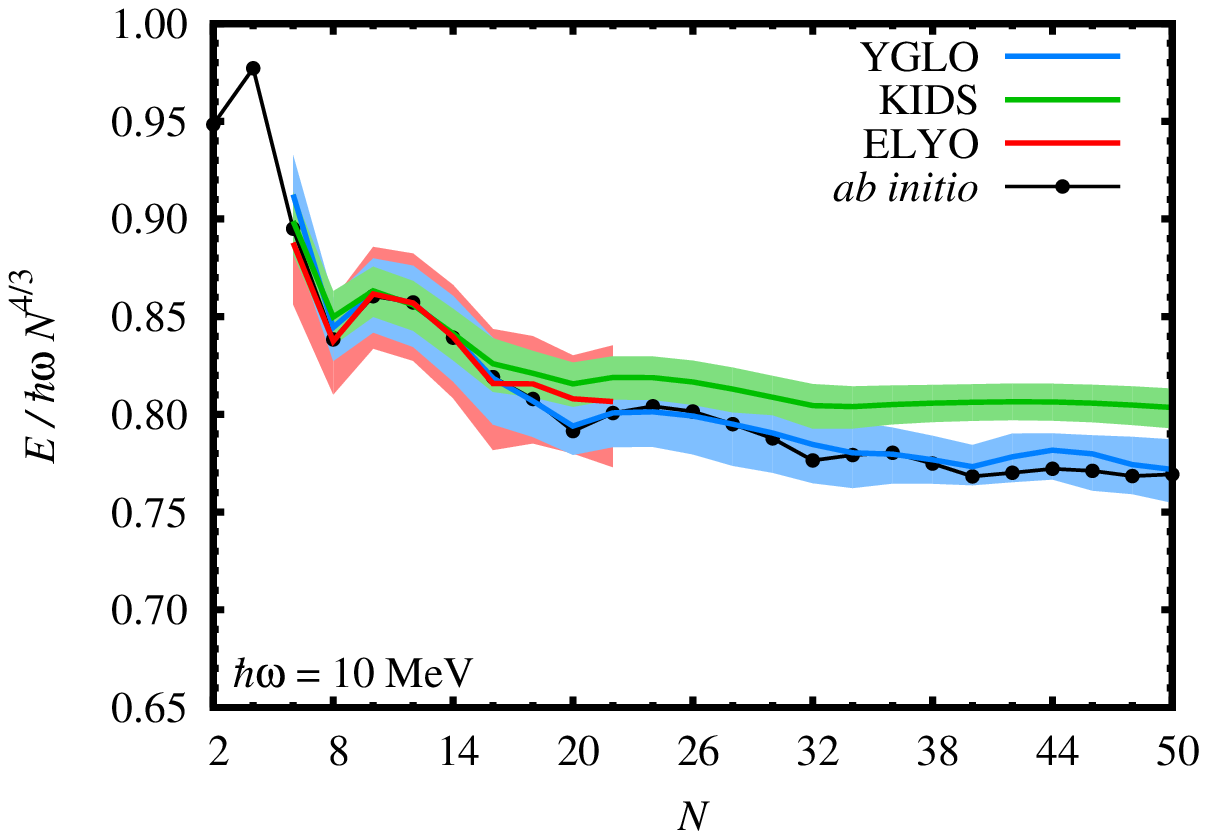}
\caption{Energies of neutron drops as a function of $N$ obtained for $\hw=10$ MeV with the YGLO, KIDS, and ELYO EDFs. The areas represent variations of $\pm 25$\% of the splitting coefficient $W$ with respect to the optimal values $W_{\rm opt}$ reported in Table \ref{tab_CCfit}. The spin--orbit and pairing couplings are kept fixed. Note that for YGLO and KIDS the upper bound for the energy is obtained for $1.25 W_{\rm opt}$ (keeping in mind that for YGLO $W_{\rm opt}$ is negative). For the ELYO functional, the upper bound is obtained for $0.75 W_{\rm opt}$.}
\label{fig_EoSDrop_W25}
\end{center}
\end{figure}

Figure \ref{fig_EoSDrop_W25} illustrates the sensitivity of the droplet energies with respect to the $W$ parameter. The areas show the evolution of the energies when $W$ varies by $\pm 25$\% around its optimal value, with fixed $V_\mathrm{so}$ and $V_\mathrm{pp}$. We observe that the YGLO and KIDS EDFs behave in a similar way, the rescaled energies being modified by approximately 2.8\%, whereas the ELYO functional seems to be more impacted as the change in the results is almost twice larger.

%
\section{Summary and conclusion}\label{SecConc}
%
%

In this work we applied three recently proposed EDFs to the description of neutron drops confined by an external 
potential. These functionals, although rather different in their forms, have in common that they are all inspired by EFTs.  While the KIDS 
EDF proposed in Ref. \cite{KIDS1} have already been employed for atomic nuclei, for the YGLO \cite{YGLO} and the ELYO \cite{ELYO} functionals, the present study represents the first attempt to treat finite systems. Guided by the KIDS strategy, a systematic protocol is implemented to interpret different density dependences of the functionals either as contributions from density--dependent effective interaction or as $t_1$--$t_2$ Skyrme--like terms.  Spin--orbit and pairing effects are explicitly accounted for to provide a realistic description of both closed-- and open--shell neutron drops. Various sets of recent {\it ab initio} calculations computed for several trap frequencies are used both to fix the extra parameters appearing for finite systems and to compare the results for values of neutron numbers and frequencies not included in the fitting procedure.   

For both YGLO and KIDS EDFs, a good agreement with {\it ab initio} results was globally achieved even at the trap frequency 
that was not used to optimize the functionals. The ELYO EDF provides results that are less easy to converge and that, in general, may differ 
from the {\it ab initio} values for systems that were not constrained in the fitting process.  It should, however, be noted that, as already emphasized, the ELYO functional has much fewer parameters compared to the two others. Following the same philosophy as in Ref. \cite{ELYO}, that is taking as a guidance the Lee-Yang expansion at low density, additional flexibility can be reached by explicitly introducing the $p$--wave contribution with a potential density dependence in the $p$--wave scattering length. Work in this direction is currently in progress.  

We also discussed properties of the functional related to the mean-field potential, pairing correlations, and effective masses in neutron systems. 
We have shown that, despite the fact that the EFT--based functionals are all adjusted to the same properties in finite and infinite neutron systems, significant differences might occur in the prediction of these properties.

We conclude by mentioning that the present work is the first step towards application to atomic nuclei for the YGLO and ELYO EDFs including superfluidity effects. 

%
%
\appendix
\section{The Skyrme functional}
The central part of the Skyrme functional may be decomposed in zero--range, effective--mass, density--dependent, tensor, and gradient (finite-range) terms as in Ref. \cite{SLY5}, or alternatively as $\E_c^\mathrm{Sk}=\sum_{i=0,3}\E_i$ with
\begin{equation}
  \E_0 = \frac{1}{4} t_0 [(2+x_0)\rho^2 - (1+2x_0)(\rho_n^2+\rho_p^2)], \\
\end{equation}
\begin{equation}
\begin{split}
  & \E_1 =  \frac{1}{8} t_1  [(2+x_1)\tau\rho - (1+2x_1)(\tau_n\rho_n+\tau_p\rho_p)] \\
  &\qquad +\frac{3}{32}t_1\{(2+x_1)(\vec{\nabla}\rho)^2-(1+2x_1)[(\vec{\nabla}\rho_n)^2 \\ 
  &\qquad +(\vec{\nabla}\rho_p)^2]\}-\frac{1}{16}t_1x_1 \vec{J}^{\,2}+\dfrac{1}{16}t_1 (\vec{J}_n^{\,2}+\vec{J}_p^{\,2}),
\end{split}
\end{equation}
\begin{equation}
\begin{split}
  & \E_2 =  \frac{1}{8} t_2 [(2+x_2)\tau\rho+(1+2x_2)(\tau_n\rho_n+\tau_p\rho_p), \\   
  &\qquad -\frac{1}{32}t_2\{(2+x_2)(\vec{\nabla}\rho)^2+(1+2x_2)[(\vec{\nabla}\rho_n)^2 \\
  &\qquad +(\vec{\nabla}\rho_p)^2]\}-\frac{1}{16}t_2x_2\vec{J}^{\,2}-\dfrac{1}{16}t_2 (\vec{J}_n^{\,2}+\vec{J}_p^{\,2}),
\end{split}
\end{equation}
\begin{equation}
   \E_3 = \frac{1}{24} t_3\rho^\alpha [(2+x_3)\rho^2 - (1+2x_3)(\rho_n^2+\rho_p^2)].
\end{equation}
$\rho_{(n,p)}$, $\tau_{(n,p)}$, $\vec{J}_{(n,p)}$ stand for the total (no index), neutron ($n$), and proton ($p$) matter, kinetic, and spin-current densities whose expression in spherical coordinates may be found in Ref. \cite{RefSk}. Historically this functional is generated by a zero--range effective interaction at leading order, i.e., as a density--dependent two--body vertex in the particle--hole channel.

The Skyrme EDF gives rise to the following EOSs:
\begin{subequations}\label{EosSkyrme}
\begin{equation}
   \dfrac{E}{A}=K_\beta(\rho)+\dfrac{3}{8} t_0\rho+ \dfrac{1}{16}t_3 \rho^{1+\alpha}  
+ \dfrac{3}{80}\left(\dfrac{3\pi^2}{2}\right)^{2/3} \Theta_s \rho^{5/3},
\end{equation}
for SNM ($\beta=1$), and 
\begin{equation}
\begin{split}
   \dfrac{E}{A}= K_\beta(\rho) &+\dfrac{1}{4} t_0(1-x_0)\rho + \dfrac{1}{24}t_3(1-x_3)\rho^{1+\alpha}\\
  & +\dfrac{3}{40}(3\pi^2)^{2/3}(\Theta_s-\Theta_v) \rho^{5/3},
\end{split}
\end{equation}
\end{subequations}
for PNM ($\beta=0$), with $\Theta_s=3t_1+t_2(5+4x_2)$ and $\Theta_v=t_1(2+x_1)+t_2(2+x_2)$. The kinetic part reads
\begin{equation}
K_\beta(\rho)= \dfrac{3}{5}\dfrac{\hbar^2}{2m}\Bigl(\dfrac{6\pi^2}{\nu}\Bigr)^{2/3}\rho^{2/3}
\end{equation}
where the degeneracy $\nu=2,4$ for $\beta=0,1$.

Isoscalar and neutron effective masses may be defined from the $\tau$-dependent part of the functional for, respectively, SNM, and PNM:
\begin{equation} \label{effMass}
\begin{split}
&\left(\dfrac{m^*}{m}\right)^{-1}_s = 1 + \dfrac{m}{8\hbar^2}\Theta_s \rho,\\
&\left(\dfrac{m^*}{m}\right)^{-1}_n = 1 + \dfrac{m}{4\hbar^2}(\Theta_s -\Theta_v) \rho.
\end{split}
\end{equation}
%
\section*{Acknowledgment}

We would like to thank P. Papakonstantinou for useful discussions on the KIDS functional. 
This project has received funding from the European Union's Horizon 2020 research and innovation programme under Grant Agreement No. 654002.

\end{document}